\documentclass[aps,pra,amssymb,amsmath,onecolumn,superscriptaddress,showpacs]{revtex4-1}

\usepackage[latin9]{inputenc}
\usepackage{amsmath}
\usepackage{amssymb}
\usepackage{graphicx}
\usepackage{verbatim}
\usepackage{bm}
\usepackage{color}

\usepackage{graphicx,epsfig,amsfonts,amssymb}
\usepackage{bm}
\usepackage{times}
\usepackage{verbatim}
\usepackage{multirow}
\usepackage{array} 

\newcommand{\nn}{\nonumber}

\newcommand{\be}{\begin{eqnarray}}
\newcommand{\ee}{\end{eqnarray}}

\newcommand{\bs}{\begin{equation}\begin{split}}
\newcommand{\es}{\end{split}\end{equation}}

\date{\today}

\begin{document}

\title{No-go rules for 
multitime Landau-Zener models}

\author{Liping Wang}
\affiliation{ School of Physics and Electronics, Hunan University, Changsha 410082, China}
\author{Chen Sun}
\email{chensun@hnu.edu.cn}
\affiliation{ School of Physics and Electronics, Hunan University, Changsha 410082, China}

\begin{abstract}
Multitime Landau-Zener (MTLZ) model is a class of exactly solvable quantum many-body models which is multitstate and multitime generalization of the two-state Landau-Zener model. 
Currently discovered MTLZ models include the ``hypercubes'', the ``fans'' and their direct product models. In this work, we prove 
two no-go rules, named the ``no  $K_{3,3}$'' rule and the ``no $1221$'' rule, which forbid the existence of exact solutions for models with certain structures of interactions. We further apply these rules to show that for models with no more than $9$ states, besides the models mentioned above there are no other MTLZ models. We also propose a scheme to systematically classify cases that could possibly host MTLZ models. Our work could serve as a guideline to search for new exactly solvable models within the MTLZ class.
\end{abstract}

\maketitle

\section{Introduction}

In theoretical studies of quantum many-body problems, exactly solvable models are of great importance both from physical and from mathematical points of view \cite{Sutherland-2004}. Through their exact solutions, physical pictures of these models can often be understood more completely than other models; besides, they can serve as reference points for considering other related unsolvable models. Mathematically, existence of exact solutions usually implies some underlying symmetries or algebras. The search of exactly solvable quantum many-body models has been of long-term interest.

One famous exactly solvable quantum model is the Landau-Zener (LZ) model discovered in the 1930s \cite{landau,zener,majorana,stuckelberg}, which is a two-state model whose Hamiltonian depends linearly on time. Multistate generalizations of the LZ model have later been studied and different types of models were found to be exactly solvable \cite{DO,Hioe-1987,bow-tie,GBT-Demkov-2000,GBT-Demkov-2001,chain-2002,4-state-2002,4-state-2015,6-state-2015,DTCM-2016,DTCM-2016-2,quest-2017,HC-2017,large-class},
in the sense that transition probabilities between asymptotic states for an evolution from $t=-\infty$ to $t=\infty$ according to the Schr\"odinger equation can be obtained analytically in these models. Solutions of these models were either derived by detours on complex plane, Laplacian transformations, usage of special functions or Lie algebraic techeniques, or obtained by applying the independent crossing approximation \cite{B-E-1993} with confirmations by numerics. Applicability of these methods always requires strict constraints on parameters of the models, and solution to a general multistate LZ model is not known. In \cite{Patra-2015} it was conjectured that special commutation relations lead to solvable multistate LZ models. The recent discovery of ``integrability conditions'' for time-dependent quantum Hamiltonians \cite{commute} points out a new method to find new solvable time-dependent quantum models. Models that satisfy the integrability conditions may be exactly solvable, and these conditions applied to Hamiltonians with certain structures give a set of constraints on the Hamiltonian's parameters, solving which will identify models that could be solvable. This idea has been applied to different types of Hamiltonians, and many new solvable models have been found \cite{Yuzbashyan-2018,MTLZ,parallel-2020,quadratic-2021}.

This work focuses on a special class of multistate LZ models that satisfy the integrability conditions, called the multitime LZ (MTLZ) model, which is multitstate and multitime generalization of the two-state LZ model (definition of the model will be presented in Section II). This class of models was introduced in \cite{large-class} and analyzed in great detail in \cite{MTLZ}.  In \cite{large-class}, it was proved that any MTLZ model can be exactly solved. In \cite{MTLZ}, an approach to classify such solvable models was developed with the help of presenting the parameters of the Hamiltinian as data on graphs. The integrability conditions lead to constraints so that only certain kinds of graphs may support solutions. In \cite{MTLZ}, the ``hypercube'' and the ``fan'' graphs were identified to support MTLZ models; several other graphs with not large numbers of vertices were analyzed but proved to have no solutions. It was thus conjectured in \cite{MTLZ} that the hypercubes, the fans and graphs corresponding to direct products of these models are the only graphs that support MTLZ models. 
However, as the number of possible graphs increases drastically with the number of vertices, complete analysis on all possible graphs with increasing number of vertices seems to require tremendous amounts of 
work. 

In this work, we prove 
two ``no-go rules'' which say that graphs containing certain kinds of structures do not support MTLZ models. 
These rules significantly reduce the number of possible graphs that need be considered. We also present a method to systematically classify graphs that may support solutions, which could serve as a guideline to find new MTLZ models, or otherwise, as a way towards an ultimate proof of the conjecture that there are no MTLZ models other than the hypercubes, the fans and their direct products.

This paper is organized as follows. Section II serves as an introduction to the MTLZ model and its graph representation. In Section III, we prove the two no-go rules which strongly restrict structures of graphs that could host MTLZ models. In Section IV, we explore consequences of the no-go rules by analyzing possible graphs with given numbers of states, and show that for graphs with no more than $9$ states, there are no other MTLZ models besides the square, the cube and the fans. We also propose a scheme to systematically classify graphs that could possibly host solvable models. Conclusions are presented in Section V.

\section{Basics for multitime Landau-Zener model}
Before presenting the no-go rules, we will give an introduction to the MTLZ model and its graph representation. It is a brief summary of the results in \cite{MTLZ} (citation to \cite{MTLZ} will not be specified in later parts of this section), and the readers interested in the whole story are referred to \cite{MTLZ}.

\subsection{Definition of the model}
The multistate Landau-Zener (LZ) model is described by a Schr\"odinger equation with a linearly time-dependent Hamiltonian:
\begin{equation}
i\frac{d}{dt}\psi={H}(t)\psi,\quad  {H}(t) = {A} +{B}t,
\label{multistate LZ}
\end{equation}
where ${A}$ and ${B}$ are Hermitian $N\times N$ matrices, 
$\psi$ is a vector with $N$ components, and $\hbar$ is set to $1$. At $N=2$, it is the two-state LZ model  
\cite{landau,zener,majorana,stuckelberg}, which is exactly solvable (namely, transition probabilities between asymptotic states for the evolution from $t=-\infty$ to $t=\infty$ can be obtained analytically), whereas at a larger $N$, solution of the model (\ref{multistate LZ}) with general choice of parameters hasn't been found. However, according to \cite{commute}, a multistate LZ model may be solvable if its Hamiltonian in \eqref{multistate LZ} is one of a family of Hamiltonians $H_j$ ($j=1, \ldots, M$, $M>1$), and this family satisfies certain ``integrability conditions''. These conditions are derived by requiring consistency of the ``multitime Schr\"odinger equations'':
\begin{equation}
\label{system1}
 i \partial \psi(\bm{x})/\partial  x^j  = {H}_{j} (\bm{x}) \psi(\bm{x}), \; \phantom{\sum} j =  1, \ldots, M, \quad M>1,
\end{equation}
where the vector $\psi(\bm{x})$ and the Hamiltonians $H_j$ depend on the ``time-vector'' $ {\bm{x}} =(x^1,\ldots, x^M)$. The parameter $x^1$ can be identified with the physical time $t$, and $H_{1}$ the multistate LZ Hamiltonian $H(t)$ in \eqref{multistate LZ}. For real matrices $H_j$ the integrability conditions are:
\be
[{H}_i,{H}_j]=0,
\label{cond1}
\ee
\be
\partial {H}_i/\partial x^j =  \partial {H}_j/\partial x^i,  \quad i,j=1,\ldots, M.
\label{cond2}
\ee
To obtain 
a multitime Landau-Zener (MTLZ) model, we further require the Hamiltonians $H_j(\bm{x})$ to be linear in $ {\bm{x}} =(x^1,\ldots, x^M)$, namely,
\begin{eqnarray}
\label{linear-family} H_{j} (\bm{x}) = B_{kj} x^{k} + A_{j}, \quad j,k=1,\ldots, M,
\end{eqnarray}
where $B_{kj}$, $A_j$ are real symmetric matrices and summations over repeated upper and lower indices are assumed. Note that, with the linear Hamiltonians (\ref{linear-family}), each of the equations in (\ref{system1}) can be viewed as an  multistate LZ  model of the form (\ref{multistate LZ}) if we identify $x^j$ with $t$. It is for this reason that the system of equations (\ref{system1}) with the set of Hamiltonians of the form (\ref{linear-family}) was named the {\it multitime} Landau-Zener model.

In \cite{large-class}, it was proved that any multistate LZ model that can be generated from such a family (\ref{linear-family}) can be exactly solved. In \cite{MTLZ}, an approach to classify such solvable models was developed with the help of presenting the parameters as data on graphs, which we are going to describe in the next subsection.

\subsection{MTLZ families on graphs}
\label{sec:integr-cond-LF-LA}

Substitutions of Eq.~\eqref{linear-family} into (\ref{cond1}) and (\ref{cond2}) give matrix  relations  for an MTLZ family:
\begin{eqnarray}
\label{linear-family-2} && B_{kj} = B_{jk}, \;\;\; [B_{jk}, B_{lm}] = 0, \\
\label{linear-family-3} && [B_{sj}, A_{k}] - [B_{sk}, A_{j}] = 0, \\
\label{linear-family-4} &&[A_{j}, A_{k}] = 0,  \qquad k,j,l,m,s=1,\ldots M,
\end{eqnarray}
Note that the lower indices  are not indices of matrix elements but rather indices that enumerate independent Hamiltonians in an MTLZ family. We will call the number of independent Hamiltonians, $M$, the {\it dimension of the MTLZ family}. Eq.~(\ref{linear-family-2}) indicates that all matrices $B_{jk}$ can be diagonalized simultaneously in some orthonormal basis set, which we will call the {\it diabatic states}.


To a given MTLZ family we associate an  undirected graph $\Gamma = (\Gamma_{0}, \Gamma_{1})$, whose vertices $a \in \Gamma_{0}$ ($a=1,\ldots, N$) represent the diabatic basis states and  edges $ab \in \Gamma_1$ correspond to the nonzero couplings between the diabatic states. The parameters of this MTLZ family are presented by three types of data on this graph:

1. Quadratic form $\Lambda^{a}$ on a vertex $a$. Let $\Lambda_{kj}^a$, $a=1,\ldots, N$ be eigenvalues of the matrices $B_{kj}$. 
To each vertex $a$ we associate a quadratic 
form
\begin{eqnarray}
\label{Lambda-quadr-form} \Lambda^{a} = \Lambda_{jk}^{a} dx^{j} \otimes dx^{k},
\end{eqnarray}
where ``$\otimes$'' denotes the tensor direct product.

2. Linear form $A^{ab}$ on an edge $ab$. The nonzero couplings $A_{j}^{ab}$ will be considered as $j$-components of a 
linear form
\begin{eqnarray}
\label{A-1-form}
A^{ab} = A^{ba} = A_{j}^{ab} dx^{j}.
\end{eqnarray}

3. Antisymmetric parameter $\gamma^{ab}=-\gamma^{ba}\ne 0$ on an edge $ab$. We also define explicitly its sign $s^{ab}$ (namely, $\gamma^{ab}=s^{ab} |\gamma^{ab}|$).

The conditions (\ref{linear-family-2})-(\ref{linear-family-4}) can be presented conveniently in terms of these three kinds of data. First, Since  $B_{kj} = B_{jk}$ due to Eq.~(\ref{linear-family-2}), we have the symmetric property $\Lambda_{jk}^{a} = \Lambda_{kj}^{a}$. Second, the condition  (\ref{linear-family-3}) implies that the introduced data satisfy
\begin{eqnarray}
\label{constr-2-LA-3} \gamma^{ab} (\Lambda^{a} - \Lambda^{b}) = A^{ab} \otimes A^{ab}.
\end{eqnarray}
This relation means that if  $A^{ab}$'s and $\gamma^{ab}$'s are known, $\Lambda^{a}$'s are also determined (up to an additional constant). We introduce the rescaled forms
\begin{eqnarray}
\label{define-bar-A} \bar{A}^{ab} = \frac{A^{ab}}{\sqrt{|\gamma^{ab}|}},
\end{eqnarray}
so that Eq.~\eqref{constr-2-LA-3} can be written as:
\begin{eqnarray}
\label{}  \Lambda^{a} - \Lambda^{b}  = s_{ab}\bar{A}^{ab} \otimes \bar{A}^{ab}.
\end{eqnarray}
It then follows the ``cycle property'' for any cycle $a_1 a_2 \ldots a_{k-1} a_k a_1$ in the graph:
\begin{eqnarray}
\label{cycle-property-bar-A} \sum_{l=1}^k s_{a_l,a_{l+1}} \bar{A}^{a_l,a_{l+1}} \otimes \bar{A}^{a_l,a_{l+1}} = 0,
\end{eqnarray}
where we identity $a_{k+1}$ with $a_1$ that appears in the summation. Third, the condition (\ref{linear-family-4}) implies that for any two  vertices $ \, a, b \in \Gamma_{0}$ that have distance $2$ (the distance between two vertices in a graph is the length of a shortest path between the two vertices), the following ``multipath property'' is satisfied:
\begin{eqnarray}
\label{constr-3-LA-bar-A} \sum_{c \in \Gamma_{0}}^{ac,bc \in \Gamma_{1}} \sqrt{|\gamma^{ac}\gamma^{bc}|} \bar{A}^{ac} \wedge \bar{A}^{bc} = 0,
\end{eqnarray}
where ``$\wedge$'' denotes the skew symmetric tensor product (the wedge product), and the summation goes over all length-$2$ paths in the graph that connect the vertices $a$ and $b$.

To summarize, the condition (\ref{linear-family-2}) was resolved automatically by going to the diabatic basis by and requiring symmetry of $\Lambda_{jk}$, and the cycle property \eqref{cycle-property-bar-A} and the multipath property \eqref{constr-3-LA-bar-A} are equivalent to the conditions (\ref{linear-family-3}) and (\ref{linear-family-4}), respectively. Thus, for an MTLZ family, the two properties \eqref{cycle-property-bar-A} and \eqref{constr-3-LA-bar-A} are completely equivalent to the general integrability conditions \eqref{cond1} and \eqref{cond2}.

In addition to the integrability conditions, we make one more requirement for the MTLZ family being considered, what we call the {\it good family} property: for any pair of distinct edges $ac, bc \in \Gamma^{1}$ that share a vertex $c$, the forms $A^{ac}$ and $A^{bc}$ are linearly independent: $A^{ac}\wedge A^{bc}\ne 0$. This property can be shown to be equivalent to the requirement that the Hamiltonians have no triple or higher order crossings of directly coupled diabatic levels at one point (namely, they have only pairwise crossings). The multistate LZ models with simultaneous multiple diabatic level crossings are interesting on their own \cite{cross}, but they are likely derivable as limits of models with only pairwise crossings.

The two properties \eqref{cycle-property-bar-A} and \eqref{constr-3-LA-bar-A} together with the good family property put constraints on the graphs which may support solutions. In particular, a graph for an MTLZ model should have the following properties:

1. (length-$2$ path property) Any pair of edges that share a vertex belongs to at least one length-$4$ cycle. Or equivalently, if two vertices are connected by a length-2 path, there must be at least two such length-2 paths.

2. (no $3$-cycle property) The graph must not have length-$3$ cycles.


The program for how to retrieve solvable families for a given graph goes as follows. First, we check whether the above ``length-$2$ path'' and ``no $3$-cycle'' properties are satisfied. If not, then there is no solvable family for this graph. Otherwise, we take the following steps: 1) We first choose the orientations on the graph, namely, fixing the sign $s^{ab}$ on every edge $\alpha=\{a,b\}$. These signs can be conveniently represented by adding arrows on edges in the original undirected graph, so it becomes a {\it directed} graph. We draw an arrow from $a$ to $b$ if $s^{ab}=-1$, and an arrow from $b$ to $a$ if $s^{ab}=1$. 2) We further identify the solutions of Eq.~(\ref{cycle-property-bar-A}), viewed as a system of bilinear equations on the forms $\bar{A}^{\alpha}$. 
Generally,  solution of Eq.~(\ref{cycle-property-bar-A}) is not unique but rather  depends on  free parameters, which we will call {\it rapidities}. 3) Once $\bar{A}^{\alpha}$ are identified, we find the solutions of Eq.~(\ref{constr-3-LA-bar-A}), viewed as a system of bilinear equations for $\sqrt{|\gamma^{ab}|}$. 
Again, the solution may not determine all $|\gamma^{ab}|$ uniquely, so some of them become free parameters of the model. At this stage, having Eq.~(\ref{define-bar-A}), we can reconstruct couplings of the Hamiltonians, which will depend on rapidities and  $|\gamma^{ab}|$. 4) Finally, the quadratic forms $\Lambda^{a}$ associated with the vertices are  obtained with Eq.~(\ref{constr-2-LA-3}). Again, this equation may not fix all $\Lambda^{a}$. The parameters that describe this freedom also become free parameters of the MTLZ Hamiltonians.

We note that the set of relations \eqref{linear-family-2}-\eqref{linear-family-4} for MTLZ families is very similar to that for the ``type $M$ families'' \cite{Owusu-2011} which are introduced to classify families of mutually commuting operators linear in a single parameter. It would be interesting to try to apply the method used in \cite{Owusu-2011} to study MTLZ models.

\subsection{Properties of the $4$-cycle graph}
\label{sec:integr-cond-LF-LAG-4loop}
\begin{figure}[!htb]
(a)~ \scalebox{0.3}[0.3]{\includegraphics{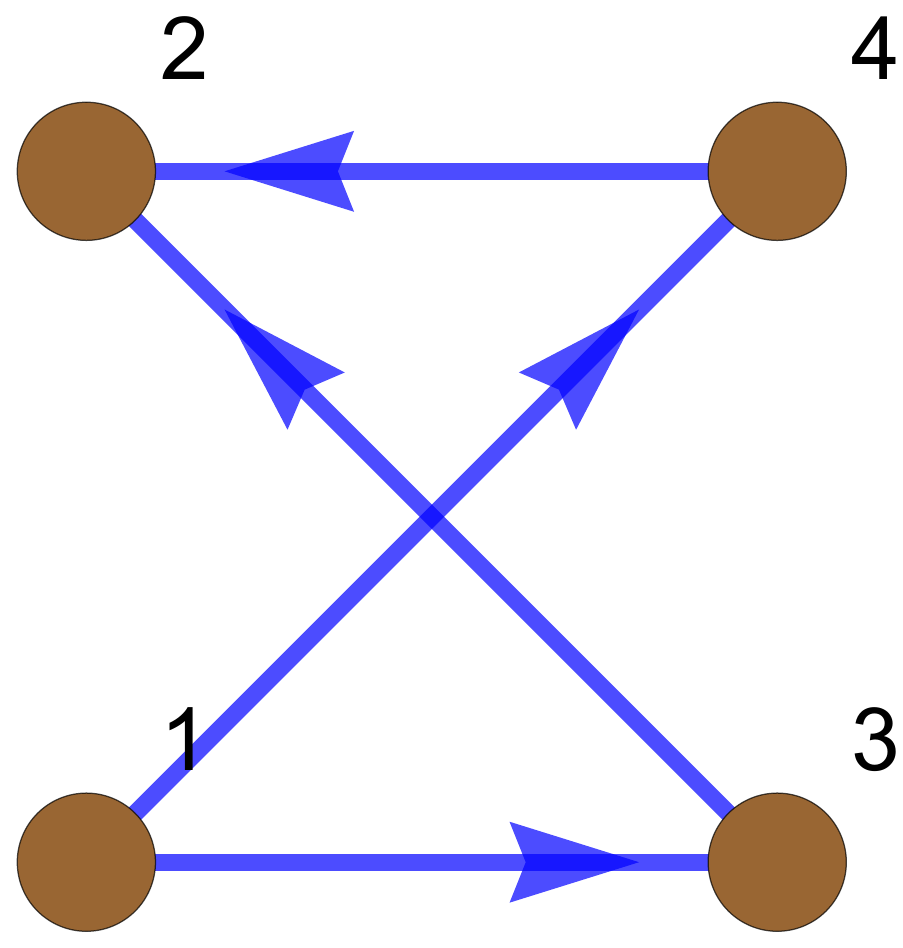}} ~ ~ ~ ~ ~ ~ ~ ~ ~ ~ ~
(b)~ \scalebox{0.3}[0.3]{\includegraphics{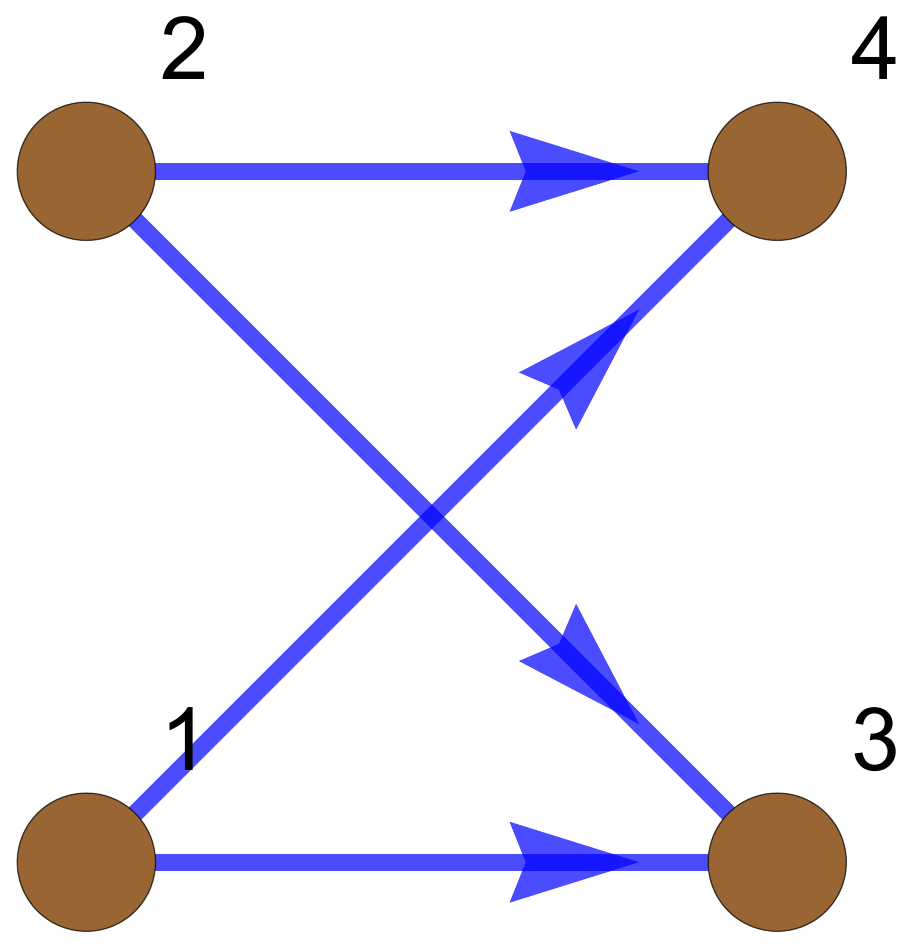}}
\caption{
Graphs of a $4$-cycle (a square, or $K_{2,2}$), with two types of orientations: (a) non-bipartite; (b) bipartite.}
\label{fig:4-loop-graph}
\end{figure}

According to the ``length-$2$ path'' and ``no $3$-cycle'' properties, the simplest and most fundamental structure in a graph for an MTLZ model is a length-$4$ cycle ($4$-cycle for short) that consists of $4$ distinct edges, as shown in Fig.~\ref{fig:4-loop-graph}. We can call this graph a ``square''. It turns out that this graph can take only two types of orientations (directions of arrows) that satisfy the cycle property \eqref{cycle-property-bar-A}, which we call the ``non-bipartite orientation'' and the ``bipartite orientation''  (note that here ``bipartite'' is different from that in ``bipartite graph'' to be introduced in Section IIIA -- here it is defined for a directed length-$4$ cycle, whereas in Section IIIA it is defined for an undirected graph. To eliminate changes of ambiguity, we will always write ``orientation'' after ``non-bipartite'' or ``bipartite'' when we are referring to directed length-$4$ cycles).




The non-bipartite orientation is shown in Fig.~\ref{fig:4-loop-graph}(a). The vertices $1$ and $2$ are a ``source'' and a ``sink'', respectively. Namely, they are the origin and the destination of all arrows that are connected to them, respectively. The other vertices $2$ and $3$ are ``intermediate'', meaning that they are neither sources or sinks. For this orientation, the $\bar A^{ab}$ forms are related by a pseudo-orthogonal transformation:
\begin{eqnarray}
\label{A-LO-transf-correspond-2} \bar{A}^{24} &=& p \left(\cosh\vartheta \bar{A}^{13} - r \sinh\vartheta \bar{A}^{14}\right), \nonumber \\ \bar{A}^{23} &=& p\left(\sinh\vartheta \bar{A}^{13} - r \cosh\vartheta \bar{A}^{14}\right),
\end{eqnarray}
where $p=\pm 1$, $r=\pm 1$, and $\vartheta$ is a free real parameter called a ``rapidity''. From this follows two relations of wedge products:
\begin{eqnarray}
\label{A-LO-determinants-tilde} \bar{A}^{23} \wedge \bar{A}^{24} = r \bar{A}^{13} \wedge \bar{A}^{14},\quad
\bar{A}^{14} \wedge \bar{A}^{24} = r \bar{A}^{13} \wedge \bar{A}^{23}.
\end{eqnarray}


The bipartite orientation is shown in Fig.~\ref{fig:4-loop-graph}(b). The vertices $1$ and $2$ are sources, and the vertices $3$ and $4$ are sinks. Again, the $\bar A^{ab}$ forms are related by a pseudo-orthogonal transformation:
\begin{eqnarray}
\label{A-LO-transf} \bar{A}^{24} &=& p \left( r \sinh\vartheta \bar{A}^{13}+\cosh\vartheta \bar{A}^{14}\right), \nonumber \\ \bar{A}^{23} &=& p\left( r \cosh\vartheta \bar{A}^{13}+\sinh\vartheta \bar{A}^{14} \right).
\end{eqnarray}
From this follows  two relations of wedge products:
\begin{eqnarray}
\label{connect-bipart} \bar{A}^{23} \wedge \bar{A}^{24} = r \bar{A}^{13} \wedge \bar{A}^{14},\quad
\bar{A}^{14} \wedge \bar{A}^{24} =- r \bar{A}^{13} \wedge \bar{A}^{23}.
\end{eqnarray}
Note that the signs for these two wedge product relations are different for the bipartite orientation, whereas those signs are the same for the non-bipartite orientation.


Derivations of the results above for the two types of orientations use only the cycle property \eqref{cycle-property-bar-A}. Thus, it applies to any $4$-cycle that appears in a graph, which may include other vertices and edges besides the considered $4$-cycle. The multipath property \eqref{constr-3-LA-bar-A}, on the other hand, requires knowledge of the entire graph, since adding new vertices and edges can bring in more paths, and thus more terms of wedge products in \eqref{constr-3-LA-bar-A}. This observation will be useful later when we derive the no-go rules.

If we now consider a $4$-cycle (Fig.~\ref{fig:4-loop-graph}) as an entire graph, then the multipath property \eqref{constr-3-LA-bar-A} gives:
\begin{eqnarray}
 \sqrt{|\gamma^{13} \gamma^{23}|}\bar{A}^{13} \wedge \bar{A}^{23} + \sqrt{|\gamma^{14} \gamma^{24}|}\bar{A}^{14} \wedge \bar{A}^{24}=0, \nn\\
 \sqrt{|\gamma^{13} \gamma^{14}|}\bar{A}^{13} \wedge \bar{A}^{14} + \sqrt{|\gamma^{23} \gamma^{24}|}\bar{A}^{23} \wedge \bar{A}^{24}=0. \label{4-loop-graph-signs}
\end{eqnarray}
These two equations can be satisfied only if $\bar{A}^{13} \wedge \bar{A}^{23} =- \bar{A}^{14} \wedge \bar{A}^{24}$ and $\bar{A}^{13} \wedge \bar{A}^{14}=-\bar{A}^{23} \wedge \bar{A}^{24}$ are simultaneously satisfied. This is possible for the non-bipartite orientation with $r=-1$, but not possible for the bipartite orientation. Thus, if a $4$-cycle graph is an entire graph, it must be of the non-bipartite orientation.

\section{no-go rules}

The ``length-$2$ path'' and the ``no $3$-cycle'' properties restrict the graphs that may support solutions. However, the number of possible graphs satisfying these two properties still increases rapidly as the number of vertices $N$ increases. (For example, see Tables 1 and 2 in the next section, which shows that the numbers of possible graphs are $4$ for $N=7$ and $14$ for $N=8$.) Therefore, complete analysis on all possible graphs with increasing number of vertices seems to require large amounts of work. In this section, we prove two ``no-go'' rules that forbid existence of certain structures that can appear in a graph that may support solutions. We will see that these rules significantly reduce the number of possible graphs that may host MTLZ models.


\begin{figure}[!htb]
(a)~ \scalebox{0.4}[0.4]{\includegraphics{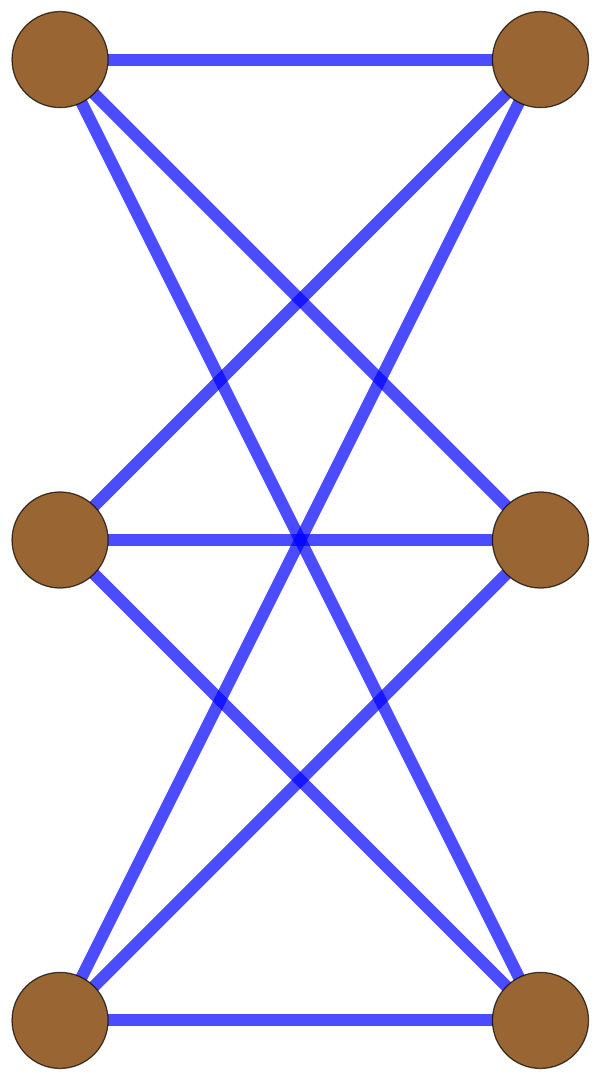}} ~ ~ ~ ~ ~ ~ ~ ~ ~ ~ ~
(b)~ \scalebox{0.6}[0.6]{\includegraphics{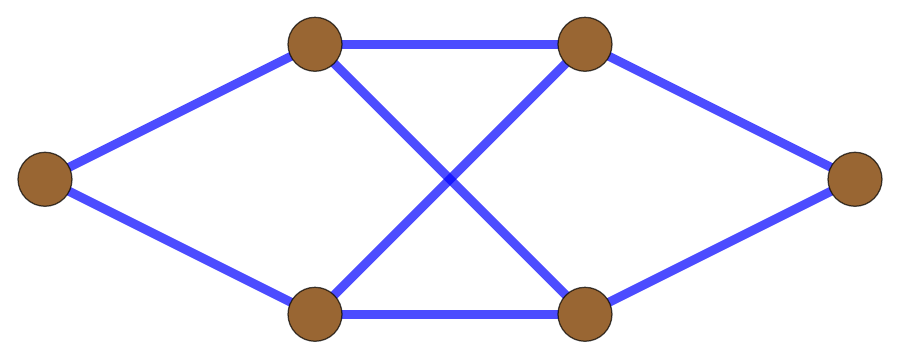}}
\caption{
The graphs appear in the statement of the no-go rules: (a) the complete bipartite graph $K_{3,3}$; (2) the layer graph ``$1221$''.}
\label{fig:K33and1221}
\end{figure}

\subsection{Statements of the no-go rules}

Let's first recall some definitions in graph theory (see, for example, \cite{Diestel}). A {\it graph} is defined as a pair of sets whose elements are vertices and edges connecting the vertices, respectively. According to this definition, two graphs are identical if they have the same number of vertices and the same structure of edges connecting the vertices. A graph is called a {\it subgraph} of another graph if the two sets of the former are subsets of those of the latter, respectively.  A graph is called a {\it bipartite graph} if its vertices can be separated into two groups such that any two vertices in the same group are not connected by an edge. A bipartite graph is said to be {\it complete bipartite} if any two vertices in the two different groups are connected by an edge. A complete bipartite graph is denoted by $K_{m,n}$, with $m$ and $n$ being the numbers of vertices in the two groups. Besides the above standard definitions in graph theory, we also introduce a ``layer graph'' notation: we call a graph a ``layer graph'' if the vertices are grouped into a sequence of ``layers'', where any vertex in a layer can be connected only to vertices in adjacent layers (i.e. not connected to vertices within the same layer and not connected to layers farther away). If such a graph is in addition maximally connected (i.e. any vertex in a layer is connected to all vertices in neighboring layers), we will name it by the series of numbers of vertices in this sequence of layers. For example, the graph shown in Fig.~\ref{fig:K33and1221}(b) will be called an ``$1221$'' graph, where the number of vertices in each layers should to be read from left to right.

For simplicity, we will call a graph ``solvable'' if it supports MTLZ models. We now state the no-go rules for solvable graphs, to be proved in the next two subsections:

\bigskip
1. {\it No $K_{3,3}$ rule}: a solvable graph must not contain $K_{3,3}$ (as shown Fig.~\ref{fig:K33and1221}(a)) as a subgraph.

\bigskip
2. {\it No $1221$ rule}: If a solvable graph contains an $1221$ layer graph as a subgraph (as shown Fig.~\ref{fig:K33and1221}(b)), there must be at least one more distinct length-$2$ path in the entire graph that connects the first and the third layer or the second and the fourth layer in this $1221$ subgraph.

\bigskip
The ``no $K_{3,3}$'' rule forbids the existence of any $K_{3,3}$ subgraph in a solvable graph. For example, any $K_{m,n}$ graph with $m\ge 3$ and $n\ge 3$ is not solvable, since it contains $K_{3,3}$ as a subgraph. The ``no $1221$'' rule, on the other hand, does not forbid the existence of an $1221$ layer subgraph in a graph, but it says that such a subgraph must be accompanied by at least one additional path as described in the rule. Note that such a path requires the presence of at least one more vertex in the entire graph but not in the $1221$ subgraph. In general, the no $K_{3,3}$ rule requires that a graph cannot be ``too connected'', whereas the no $1221$ rule requires that it must be ``connected enough''; they together form a ``window'' for the allowed graphs. 
As we will see in the next section, these two no-go rules turn out to be strong restrictions to the possible graphs that may hold MTLZ models.

\subsection{Proof of the ``no $K_{3,3}$'' rule}

\begin{figure}[!htb]
(a)~ \scalebox{0.4}[0.4]{\includegraphics{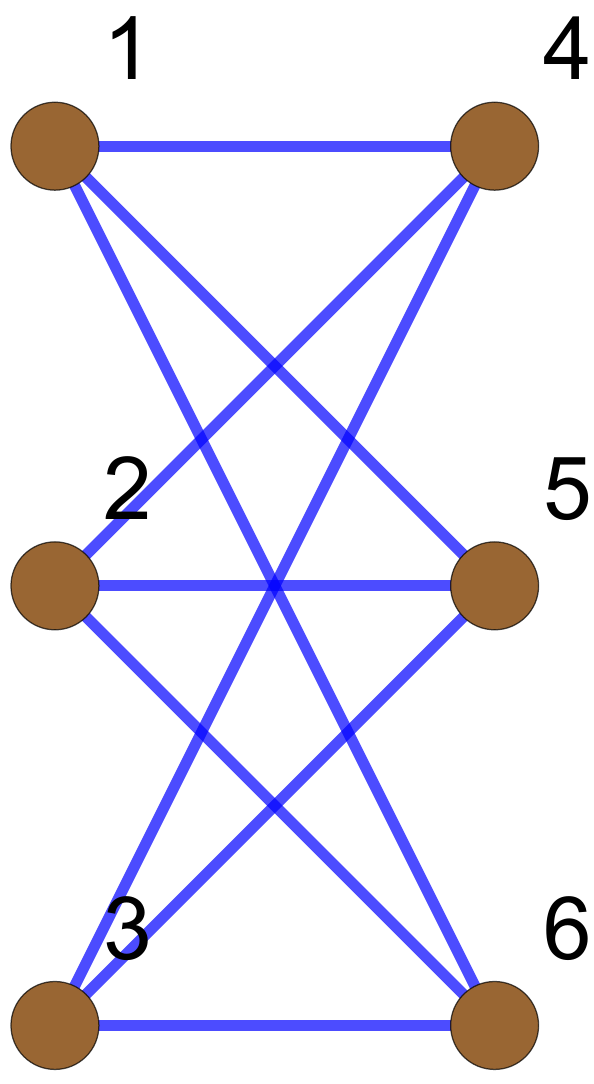}} ~ ~ ~ ~ ~ ~ ~ ~ ~ ~ ~
(b)~ \scalebox{0.4}[0.4]{\includegraphics{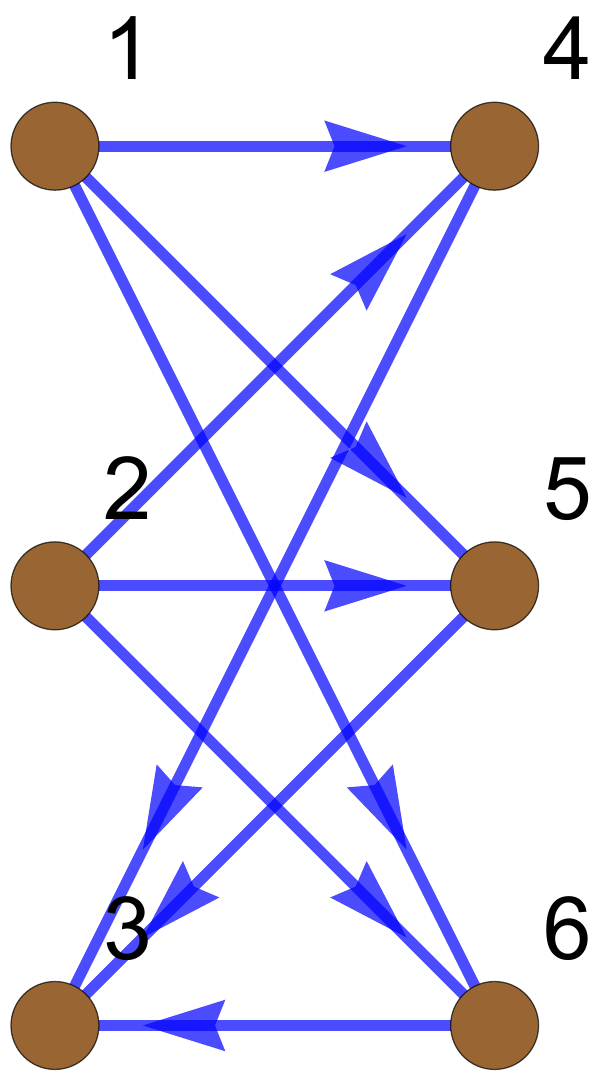}} ~ ~ ~ ~ ~ ~ ~ ~ ~ ~ ~
(c)~ \scalebox{0.4}[0.4]{\includegraphics{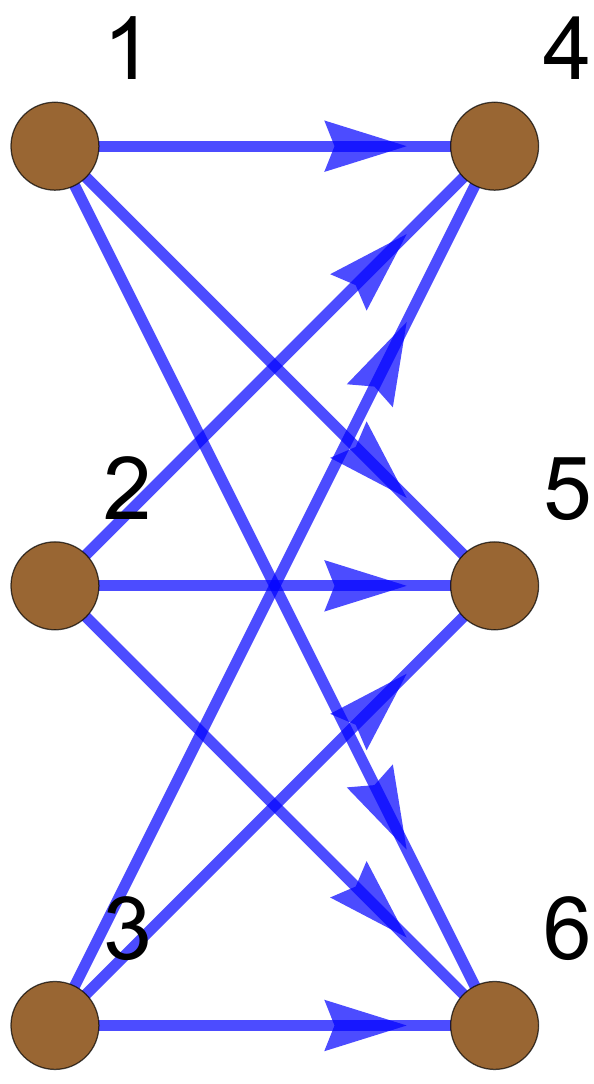}}
\caption{(a) The undirected $K_{3,3}$ graph. (b) and (c) are the possible orientations of the $K_{3,3}$ graph: (b) with $2$ sources and $1$ sink; (c)  with $3$ sources and $3$ sinks.}
\label{fig:K33}
\end{figure}

We consider a $K_{3,3}$ graph as shown in Fig.~\ref{fig:K33}(a), with the vertices labelled by numbers from $1$ to $6$. The graph contains a number of $4$-cycles. Let's look at one of them, say the $4$-cycle formed by the vertices $1$, $2$, $4$ and $5$. We will denote it as $1425$. According to the results for a $4$-cycle described in Section~IIC, the $\bar A^{ab}$ forms on its edges are related by two wedge product relations:
\begin{eqnarray}
&\bar{A}^{14} \wedge \bar{A}^{24} = r_{1425} \bar{A}^{15} \wedge \bar{A}^{25},\label{eq:wedge-1425}\\
&\bar{A}^{14} \wedge \bar{A}^{15} = r_{4152} \bar{A}^{24} \wedge \bar{A}^{25}.\label{}
\end{eqnarray}
(The relative sign of $\bar{A}^{ab} \wedge \bar{A}^{bc}$ and $\bar{A}^{ad} \wedge \bar{A}^{cd}$ is denoted as $r_{abcd}$.) The signs $ r_{1425}$ and $ r_{4152}$ are related according to the type of orientation of this $4$-cycle:
\begin{align}
&r_{1425}=r_{4152}, \quad \textrm{if the cycle $1425$ is of the non-bipartite orientation,} \label{eq:rnb}\\
&r_{1425}=-r_{4152}, \quad \textrm{if the cycle $1425$ is of the bipartite orientation.} \label{eq:rb}
\end{align}
Similarly, for the cycle $1426$, we have
\begin{eqnarray}
&\bar{A}^{14} \wedge \bar{A}^{24} = r_{1426} \bar{A}^{16} \wedge \bar{A}^{26},\label{eq:wedge-1426}\\
&\bar{A}^{14} \wedge \bar{A}^{16} = r_{4162} \bar{A}^{24} \wedge \bar{A}^{26},\label{}
\end{eqnarray}
and for the cycle $1526$, we have
\begin{eqnarray}
&\bar{A}^{15} \wedge \bar{A}^{25} = r_{1526} \bar{A}^{16} \wedge \bar{A}^{26},\label{eq:wedge-1526}\\
&\bar{A}^{15} \wedge \bar{A}^{16} = r_{5162} \bar{A}^{25} \wedge \bar{A}^{26}.\label{}
\end{eqnarray}
Eqs.~\eqref{eq:wedge-1425}, \eqref{eq:wedge-1426} and \eqref{eq:wedge-1526} together imply that the three signs for the three $4$-cycles involving vertices $1$ and $2$ are related by
\begin{align}\label{eq:r12}
r_{1425}r_{1426}r_{1526}=1.
\end{align}
Due to the symmetry of the vertices $1$, $2$ and $3$, we can immediately write out the relation for the signs for the three $4$-cycles involving vertices $1$ and $3$:
\begin{align}\label{eq:r13}
r_{1435}r_{1436}r_{1536}=1,
\end{align}
and the relation for the signs for the three $4$-cycles involving vertices $2$ and $3$:
\begin{align}\label{eq:r23}
r_{2435}r_{2436}r_{2536}=1.
\end{align}
Multiplying Eqs. \eqref{eq:r12}, \eqref{eq:r13} and \eqref{eq:r23} together gives:
\begin{align}\label{eq:rleft}
 r_{left}\equiv r_{1425}r_{1426}r_{1526}r_{1435}r_{1436}r_{1536}r_{2435}r_{2436}r_{2536}=1.
\end{align}
Similar consideration on the $4$-cycles involving a pair of vertices from $4$, $5$ and $6$ gives:
\begin{align}\label{eq:rright}
 r_{right}\equiv r_{4152}r_{4153}r_{4253}r_{4162}r_{4163}r_{4263}r_{5162}r_{5163}r_{5263}=1.
\end{align}
We thus arrive at a relation:
\begin{align}\label{eq:rleftright}
 r_{left}= r_{right}=1.
\end{align}
This relation does not depend on the specific orientations of the $4$-cycles in the graph. On the other hand, each of the $9$ sign factors appearing in $r_{left}$ is related to a sign factor in $r_{right}$ by an equation similar to \eqref{eq:rnb} or \eqref{eq:rb}, depending on whether the corresponding $4$-cycle has a non-bipartite or bipartite orientation.

We now consider specific orientations of the graph. Up to the symmetry of permutation of indices and the symmetry of simultaneous reversal of directions of all arrows in a graph, there are two possible kinds of orientations, shown in Fig.~\ref{fig:K33}(b) and (c). In the case Fig.~\ref{fig:K33}(b), among the $6$ vertices there are $2$ sources and $1$ sink, and other $3$ intermediate; in the case Fig.~\ref{fig:K33}(c), among the $6$ vertices there are $3$ sources and $3$ sinks. For the $9$ $4$-cycles contained in the graph (namely, cycles $1425$, $1426$, $1526$, $1435$, $1436$, $1536$, $2435$, $2436$, $2536$), in the case Fig.~\ref{fig:K33}(b), the $3$ cycles $1425$, $1426$ and $1526$ are of the bipartite orientation and all others are of the non-bipartite orientation; in the case Fig.~\ref{fig:K33}(c), all the $9$ $4$-cycles are of the bipartite orientation. In both cases, the numbers of $4$-cycles in the bipartite orientation among the $9$ $4$-cycles are odd. This means that, of the $9$ relations between the corresponding pairs of signs appearing in $r_{left}$ (Eq.~\eqref{eq:rleft}) and $r_{right}$ (Eq.~\eqref{eq:rright}), there is always an odd number of bipartite sign relations like Eq.~\eqref{eq:rb}. This implies that
\begin{align}\label{eq:rleftright2}
 r_{left}= -r_{right}.
\end{align}
Eqs.~\eqref{eq:rleftright} and \eqref{eq:rleftright2} obviously contradict each other, so the considered $K_{3,3}$ graph does not support a MTLZ model.

Finally, we note that the above argument uses only results from the cycle property and does not use the multipath property, so it applies to any $K_{3,3}$ structure that appears in a graph. Thus, any graph that contains $K_{3,3}$ as a subgraph is not solvable. 

\subsection{Proof of the ``no $1221$'' rule}

\begin{figure}[!htb]
(a)~ \scalebox{0.6}[0.6]{\includegraphics{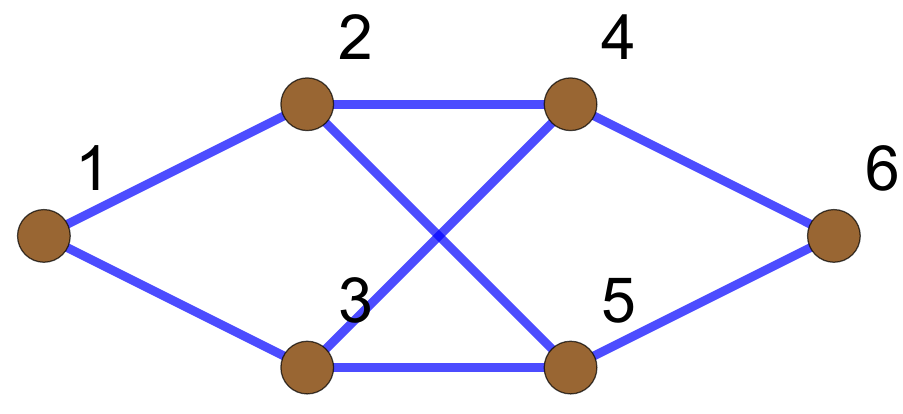}} \\
(b)~ \scalebox{0.6}[0.6]{\includegraphics{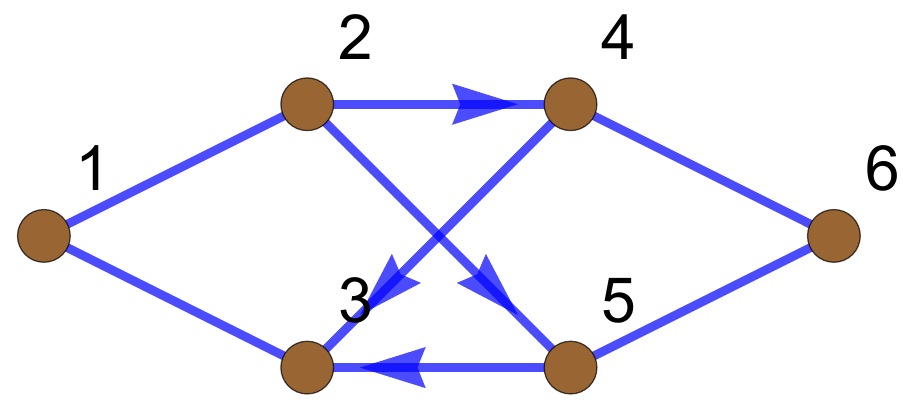}} ~ ~ ~ ~ ~ ~ ~ ~ ~ ~ ~
(c)~ \scalebox{0.6}[0.6]{\includegraphics{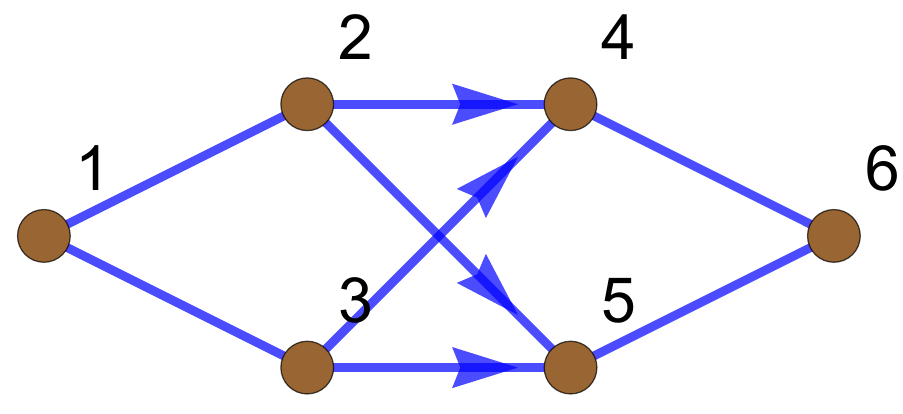}}
\caption{(a) The undirected $1221$ graph. (b) The $1221$ graph where the cycle $2435$ is of the non-bipartite orientation. (c) The $1221$ graph where the cycle $2435$ is of the bipartite orientation.}
\label{fig:1221}
\end{figure}

We consider an $1221$ graph as shown in Fig.~\ref{fig:1221}(a), with the vertices labelled by numbers from $1$ to $6$. Similar to the derivation of Eq.~\eqref{eq:r12}, the wedge product relations among $\bar{A}^{12} \wedge \bar{A}^{13}$, $\bar{A}^{24} \wedge \bar{A}^{34}$ and $\bar{A}^{25} \wedge \bar{A}^{35}$ give:
\begin{align}\label{eq:fan12345}
r_{2134}r_{2135}=r_{2435}.
\end{align}
And the wedge product relations among $\bar{A}^{24} \wedge \bar{A}^{25}$, $\bar{A}^{34} \wedge \bar{A}^{35}$ and $\bar{A}^{46} \wedge \bar{A}^{56}$ give:
\begin{align}\label{eq:fan23456}
r_{4256}r_{4356}=r_{4253}.
\end{align}
These two relations are valid for any choices of orientations of the $4$-cycles in the graph.

The $4$-cycle in the middle, namely, the cycle $2435$, can take non-bipartite or bipartite orientations as shown in Fig.~\ref{fig:1221}(b) and (c), respectively. We will consider these two cases separately.

If the cycle $2435$ is of the non-bipartite orientation (as in Fig.~\ref{fig:1221}(b)), then we have
\begin{align}\label{eq:2435-nb}
r_{2435}=r_{4253}.
\end{align}
For this case, the cycles $1243$ and $1253$ both have to be of the non-bipartite orientation, so
\begin{align}\label{}
r_{1243}=r_{2134},\quad r_{1253}=r_{2135}.
\end{align}
The orientations of the cycles $2465$ and $3465$ are not determined, but their orientations are related to each other -- if one of the two is of the non-bipartite orientation, then the other needs to be of the bipartite orientation, since in the cycle $2465$ the vertex $2$ is a source, whereas in the cycle $3465$ the vertex $3$ is a sink. We thus have the following relation for the sign factors:
\begin{align}\label{eq:4r-nb-fan23456}
r_{2465}r_{3465}=-r_{4256}r_{4356}.
\end{align}
Combining Eqs.~\eqref{eq:fan12345}--\eqref{eq:4r-nb-fan23456}, we get:
\begin{align}\label{eq:4r-nb}
r_{1243}r_{1253}r_{2465}r_{3465}=-1.
\end{align}

If the cycle $2435$ is of the bipartite orientation (as in Fig.~\ref{fig:1221}(c)), then we have
\begin{align}\label{eq:2435-b}
r_{2435}=-r_{4253}.
\end{align}
The orientations of the cycles $1243$ and $1253$ are not determined, but their orientations are related to each other -- they need to be simultaneously of the non-bipartite or the bipartite orientation, since in the cycle $1243$ the vertex $4$ is a sink, and in the cycle $3465$ the vertex $3$ is also a sink. We thus have:
\begin{align}\label{}
r_{1243}r_{1253}=r_{2134} r_{2135}.
\end{align}
Similarly, the orientations of the cycles $2465$ and $3465$ are not determined, but they need to be simultaneously of the non-bipartite or the bipartite orientation, since in the cycle $2465$ the vertex $2$ is a source, and in the cycle $3465$ the vertex $3$ is also a source. We thus have:
\begin{align}\label{eq:4r-b-fan23456}
r_{2465}r_{3465}=r_{4256}r_{4356}.
\end{align}
Combining Eqs.~\eqref{eq:fan12345}, \eqref{eq:fan23456} and \eqref{eq:2435-b}--\eqref{eq:4r-b-fan23456}, we get:
\begin{align}\label{eq:4r-b}
r_{1243}r_{1253}r_{2465}r_{3465}=-1.
\end{align}

Thus, we see that no matter the cycle $2435$ is of the non-bipartite orientation or of the bipartite orientation, the relation
\begin{align}\label{eq:4r}
r_{1243}r_{1253}r_{2465}r_{3465}=-1
\end{align}
is always satisfied. It's important to note that the above analysis uses only the cycle property, so it applies to any $1221$ structure that appears in a graph. In other words, for any $1221$ graph as a subgraph of a larger solvable graph, Eq.~\eqref{eq:4r} always holds.

We now make use of the multipath property. Let's assume that there is no more length-$2$ paths between the vertices $1$ and $4$ except the two paths $124$ and $134$, then the multipath property requires that:
\begin{eqnarray}
\label{}  \sqrt{|\gamma^{12}\gamma^{24}|} \bar{A}^{12} \wedge \bar{A}^{24} + \sqrt{|\gamma^{13}\gamma^{34}|} \bar{A}^{13} \wedge \bar{A}^{34}= 0,
\end{eqnarray}
which can be satisfied only if
\begin{align}\label{}
r_{1243}=-1.
\end{align}
Similarly, if we also assume that there are no more length-$2$ paths between the vertices $1$ and $5$, between the vertices $2$ and $6$, and between the vertices $3$ and $6$, we will have
\begin{align}\label{}
r_{1253}=-1,\quad r_{2465}=-1,\quad r_{3465}=-1.
\end{align}
The if there is no more length-$2$ paths between the vertices $1$ and $4$, between $1$ and $5$, between $2$ and $6$, and between $3$ and $6$, we will have
\begin{align}\label{}
r_{1243}r_{1253}r_{2465}r_{3465}=1,
\end{align}
which contradicts the previously proved relation \eqref{eq:4r}. Since \eqref{eq:4r} is supposed to hold for any $1221$ subgraph in a solvable graph, this means that the assumptions above must not be all true. To satisfy  \eqref{eq:4r}, at least one signs of $r_{1243}$, $r_{1253}$, $r_{2465}$ and $r_{3465}$ need to be $1$, which means that the corresponding two vertices need to have at least one more length-$2$ path besides the two paths appeared in the $1221$ layer graph. Therefore, in the entire graph that contains this $1221$ layer graph, there must be at least one more distinct length-$2$ path that connects the vertices $1$ and $4$, or  $1$ and $5$, or $2$ and $6$, or $3$ and $6$.

\section{Application of the no-go rules}

In this section, we are going to apply the no-go rules to graphs with given numbers of states. We will see that they put very strong restrictions on structures of graphs that could host MTLZ models. In particular, we will show that for graphs with no more than $9$ states, there are no other MTLZ models besides the square, the cube and the fans, with only one possible exception. We will also propose a scheme to systematically classify graphs that could possibly host MTLZ models.

\subsection{Graphs with $N\le 8$ vertices}

\begin{table}[]\label{}
\caption{All possible connected graphs with $N\le 7$ vertices that satisfy the ``length-$2$ path'' and the ``no $3$-cycle'' properties. Names of these graphs follow from the notations introduced at the beginning of Section IIIA for complete bipartite graphs $K_{m,n}$ and for layer graphs.}
\smallskip
\begin{tabular}{cc}
  \hline
  \hline
   &  \\
  $N=2$ &  $K_{1,1}$: \scalebox{0.15}[0.15]{\includegraphics{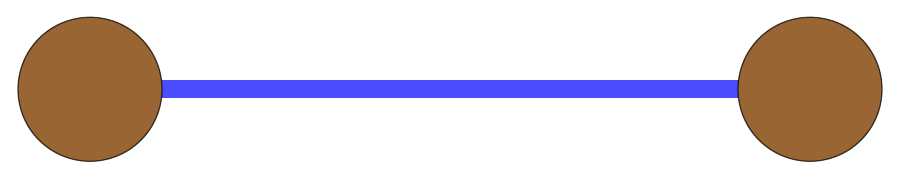}}    \\
     &  \\
  \hline
  $N=4$ &  $K_{2,2}$: \scalebox{0.15}[0.15]{\includegraphics{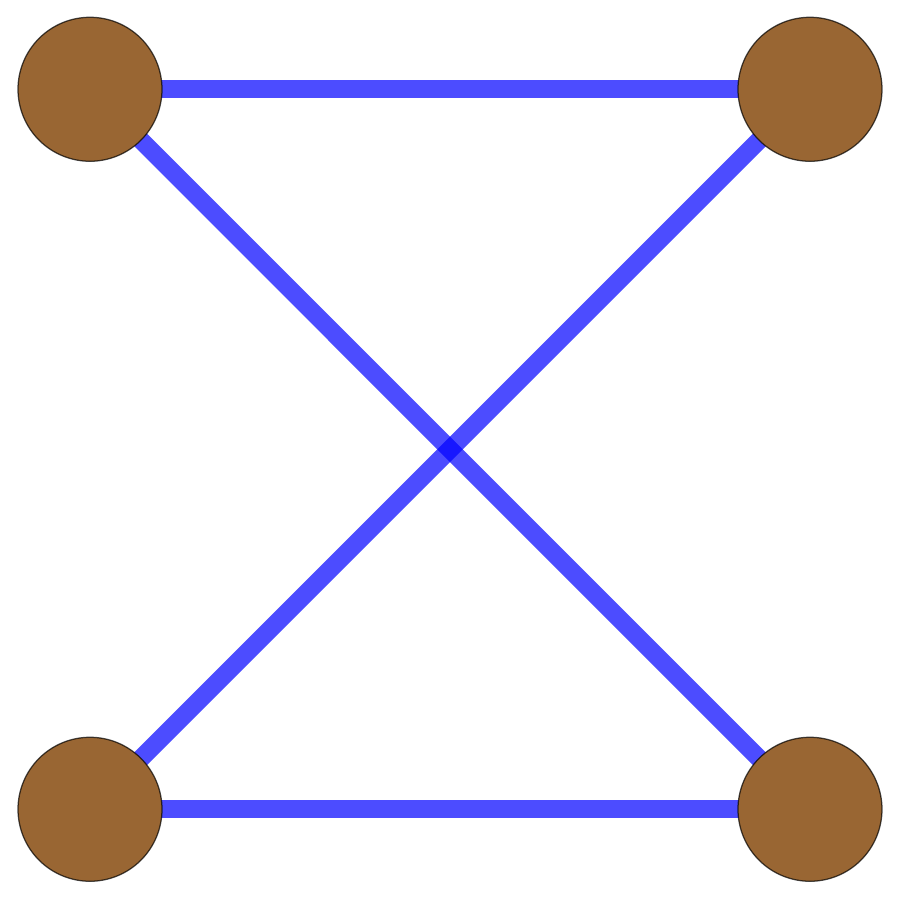}}    \\
   \hline
  $N=5$ & $K_{2,3}$: \scalebox{0.22}[0.22]{\includegraphics{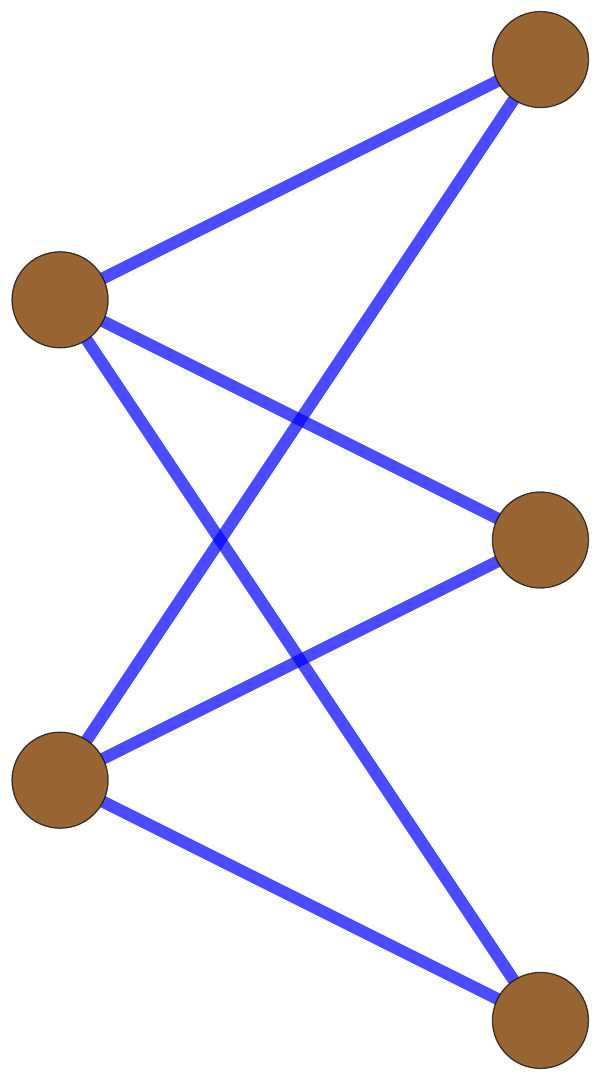}}  \\
  \hline
  $N=6$ & $K_{2,4}$: \scalebox{0.35}[0.35]{\includegraphics{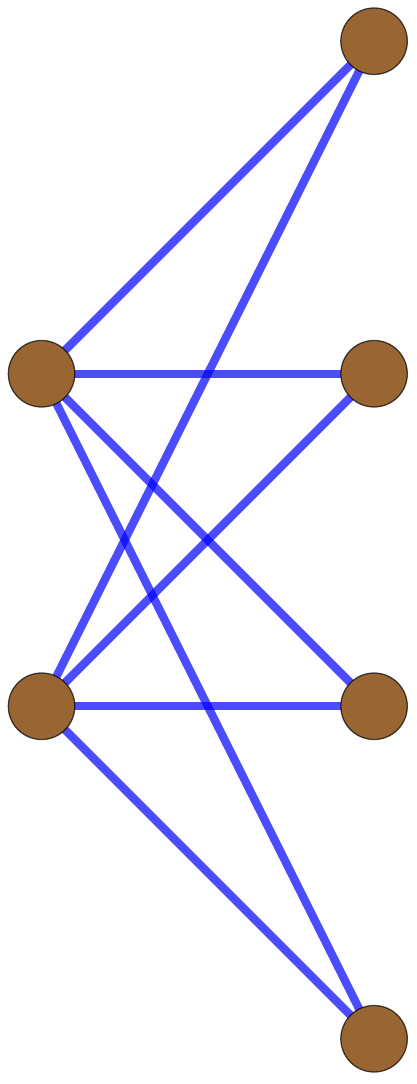}}   \qquad
$K_{3,3}$:        \scalebox{0.25}[0.25]{\includegraphics{6-3}} \qquad
  1221:   \scalebox{0.4}[0.4]{\includegraphics{6-2}}   \\
     \hline
 $N=7$ & \qquad $K_{2,5}$: \scalebox{0.4}[0.4]{\includegraphics{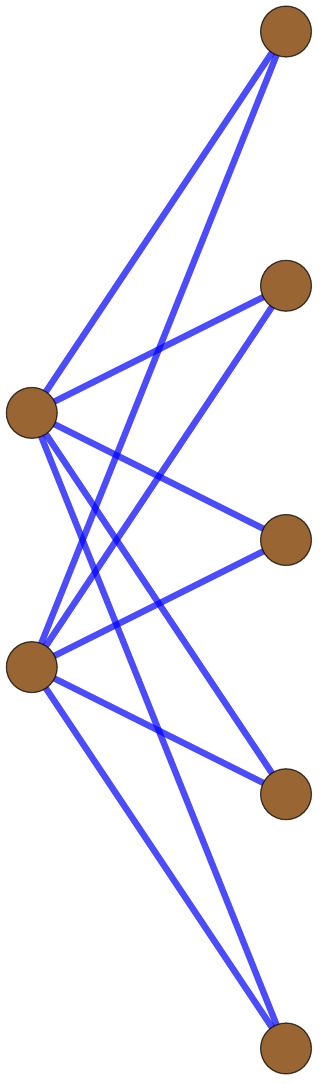}} \qquad
  $K_{3,4}$:  \scalebox{0.3}[0.3]{\includegraphics{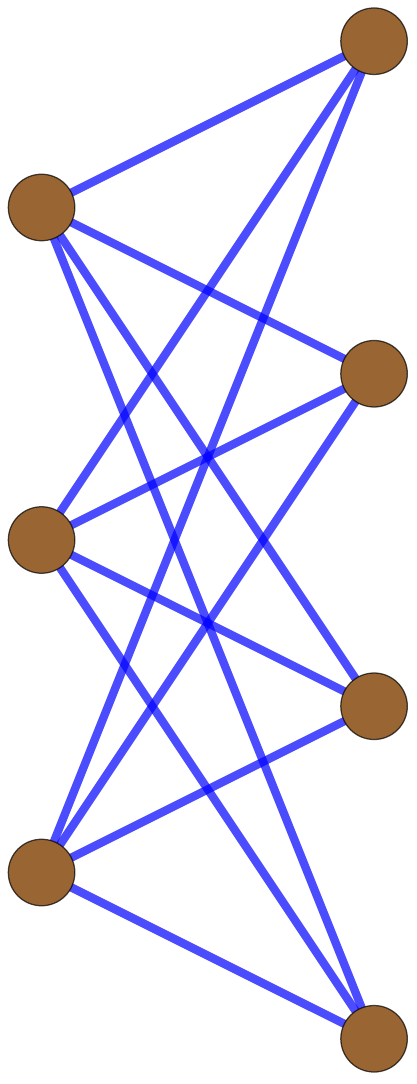}} \qquad
 1222:             \scalebox{0.4}[0.4]{\includegraphics{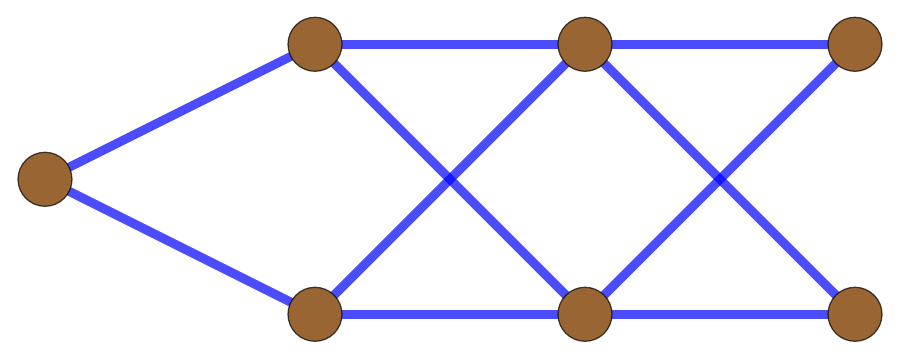}}\qquad
  1231:  \scalebox{0.4}[0.4]{\includegraphics{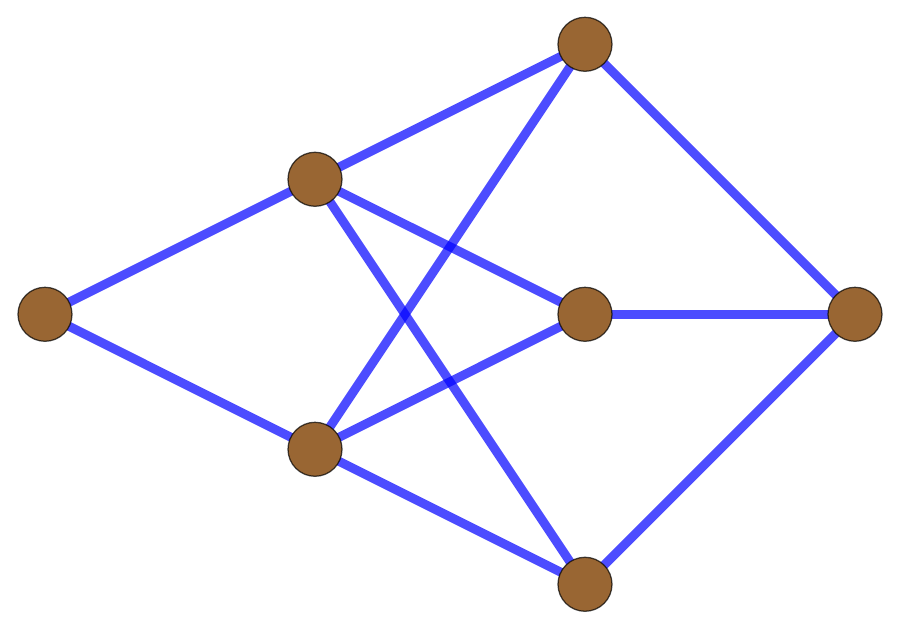}}     \\
     \hline
       \hline
\end{tabular}
\end{table}

\begin{table}[]\label{}
\caption{All possible connected graphs with $N= 8$ vertices that satisfy the ``length-$2$ path'' and the ``no $3$-cycle'' properties. Names of these graphs follow from the notations introduced at the beginning of Section IIIA for complete bipartite graphs $K_{m,n}$ and for layer graphs (except that the cube, cube$+1$, cube$+2$, and cube$+3$ graphs are a $3$-dimensional cube and cubes with one, two and three edges  added, respectively; the $1232-1$ graph is the $1232$ graph with one edge removed). }
\smallskip
\begin{tabular}{c}
  \hline
  \hline
  $K_{2,6}$: \scalebox{0.5}[0.5]{\includegraphics{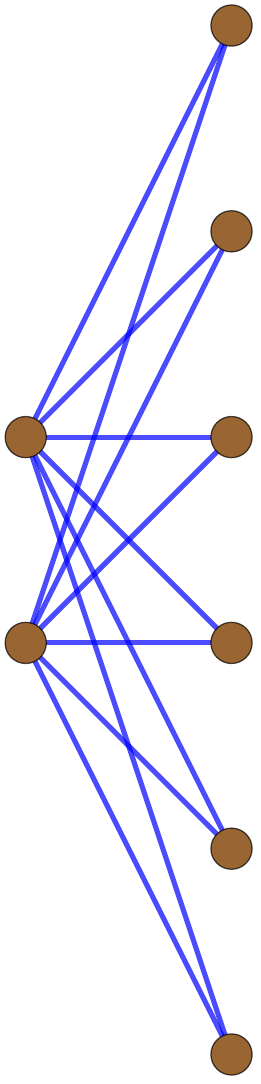}} \qquad
  $K_{3,5}$: \scalebox{0.4}[0.4]{\includegraphics{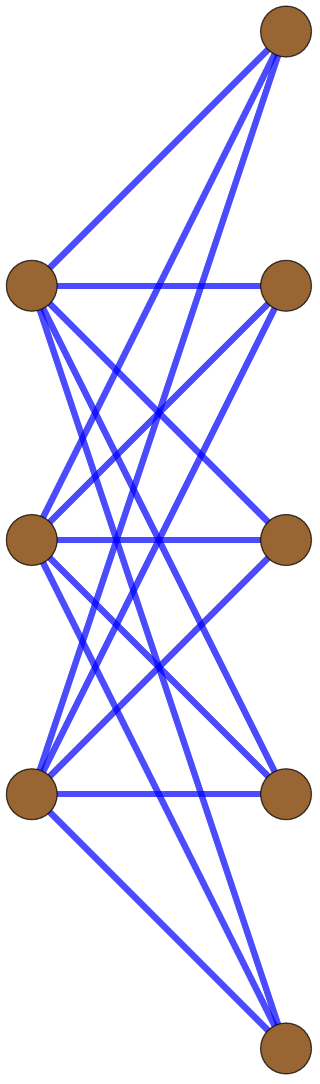}} \qquad
   $K_{4,4}$: \scalebox{0.3}[0.3]{\includegraphics{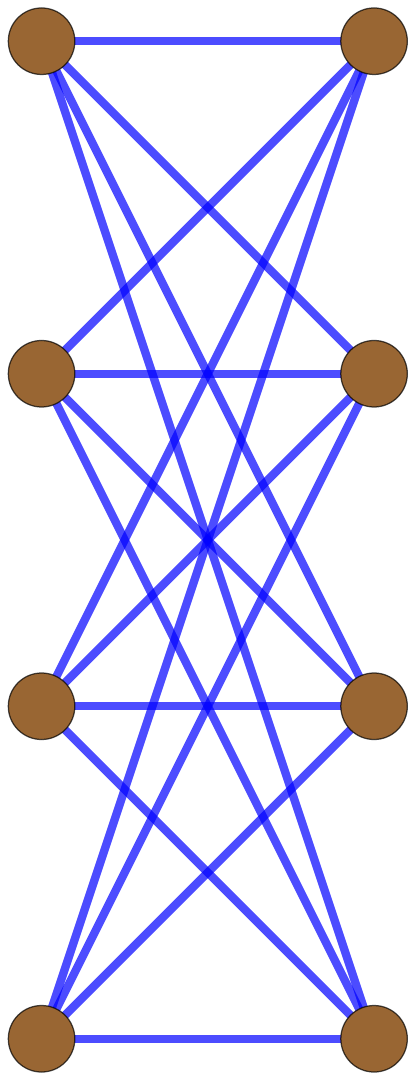}}  \qquad
     cube: \scalebox{0.4}[0.4]{\includegraphics{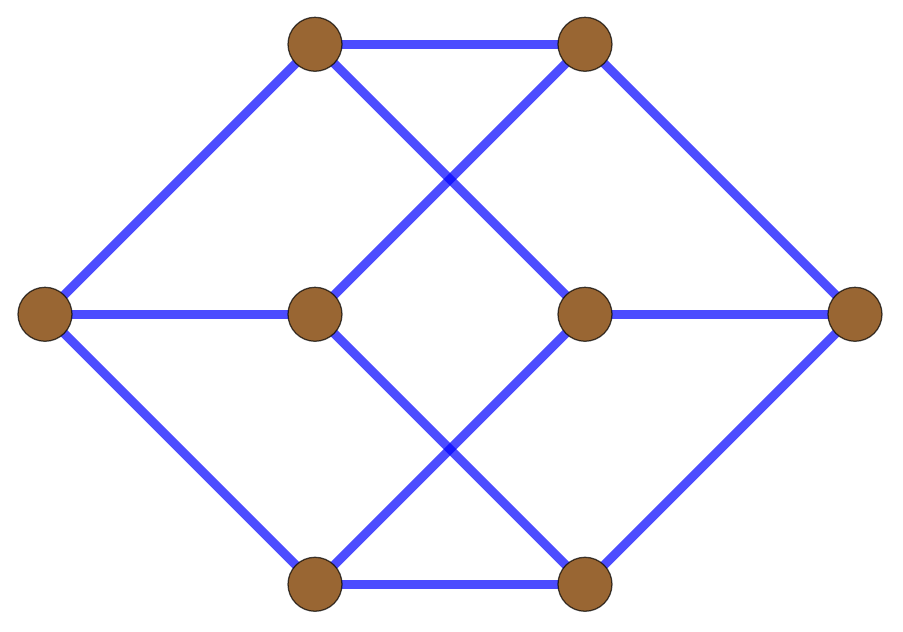}}
           \\
   \hline
    cube$+1$: \scalebox{0.4}[0.4]{\includegraphics{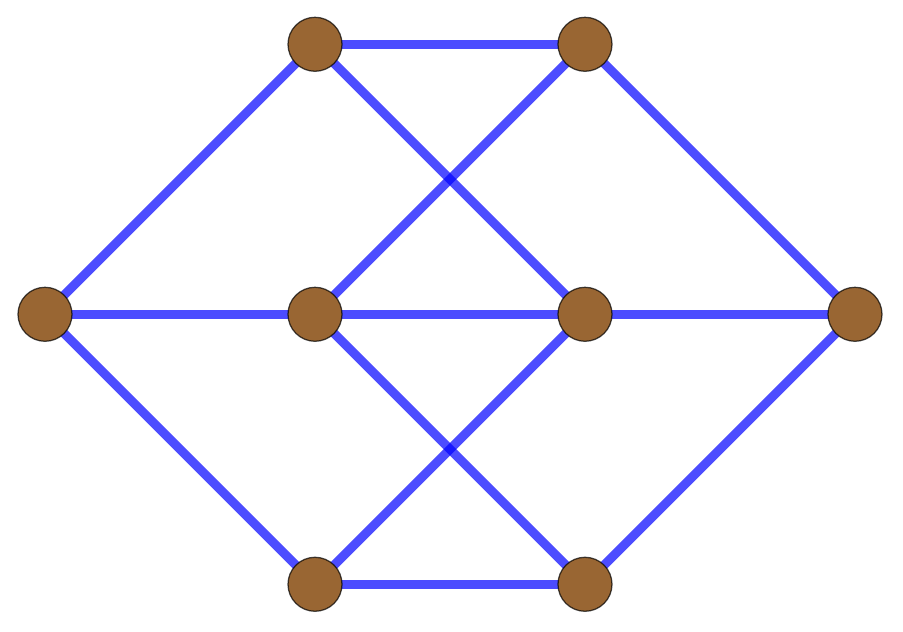}} \qquad
   cube$+2$: \scalebox{0.4}[0.4]{\includegraphics{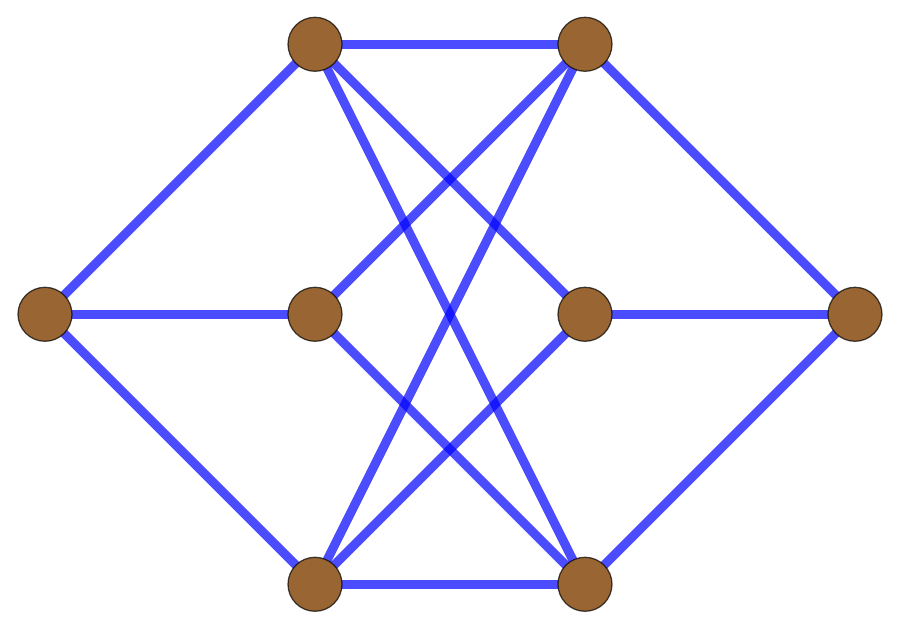}} \qquad
    cube$+3$: \scalebox{0.4}[0.4]{\includegraphics{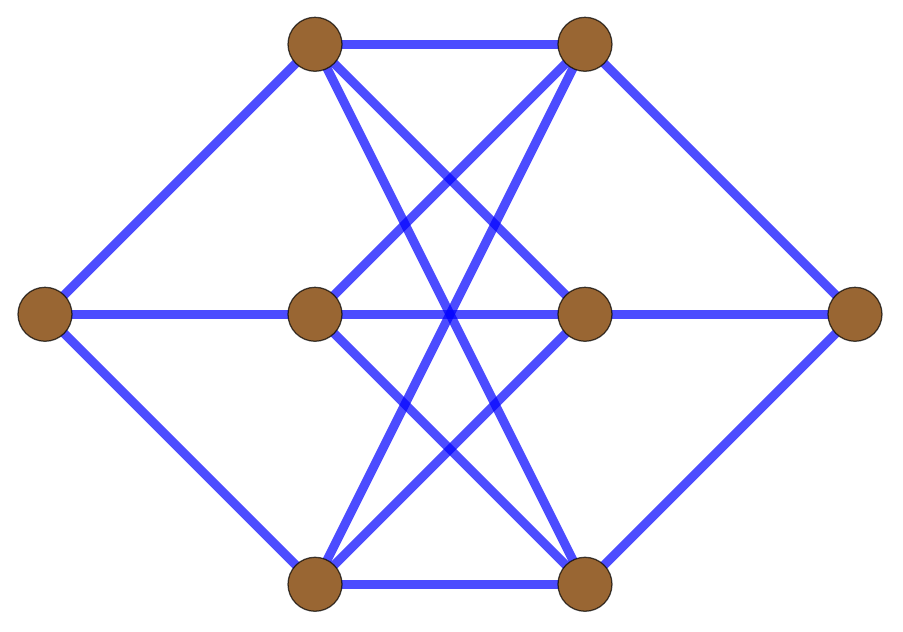}} \\
  \hline
     $1223$: \scalebox{0.4}[0.4]{\includegraphics{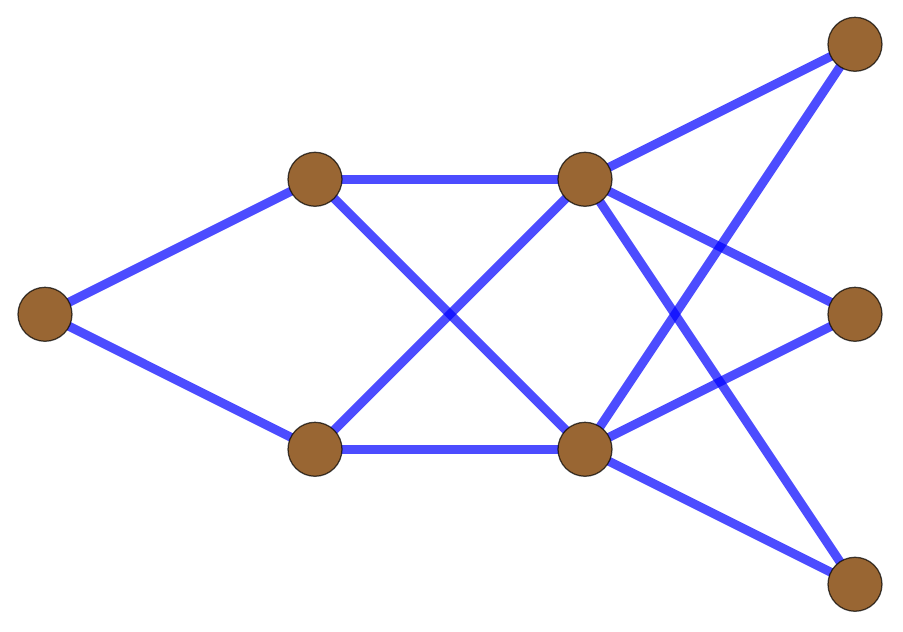}} \qquad
   $1232$: \scalebox{0.4}[0.4]{\includegraphics{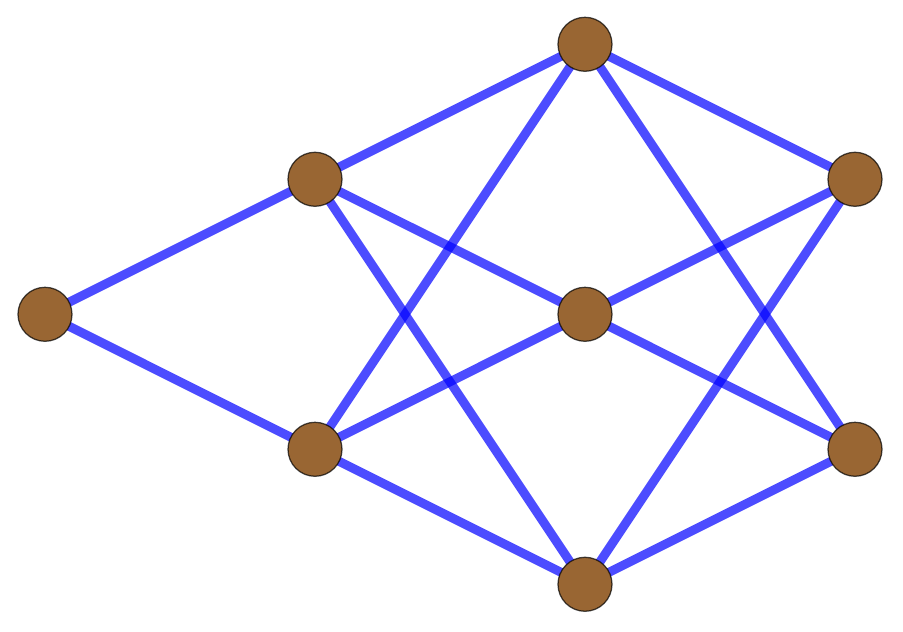}} \qquad
     $1232$$-1$: \scalebox{0.4}[0.4]{\includegraphics{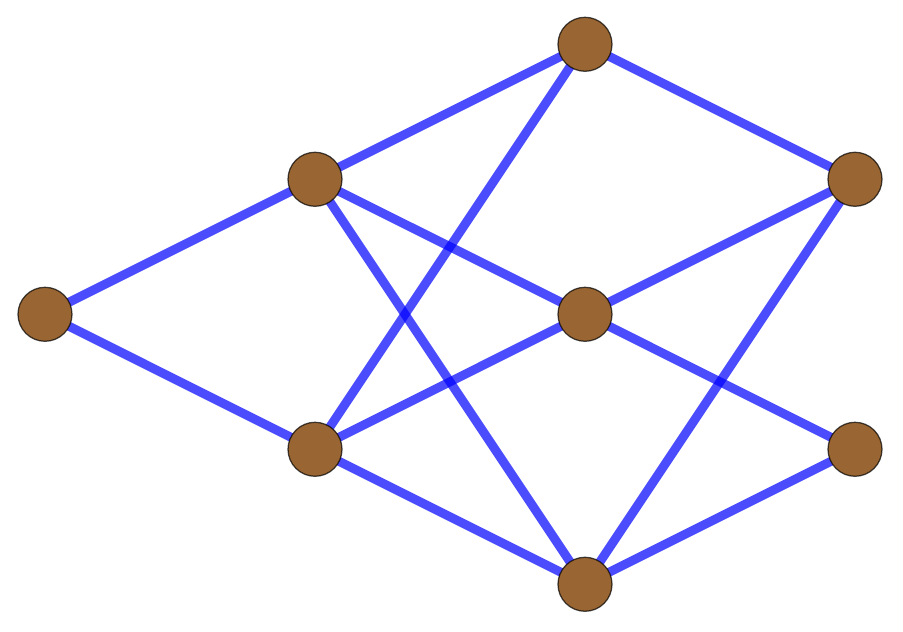}} \\
  \hline
      $1322$: \scalebox{0.4}[0.4]{\includegraphics{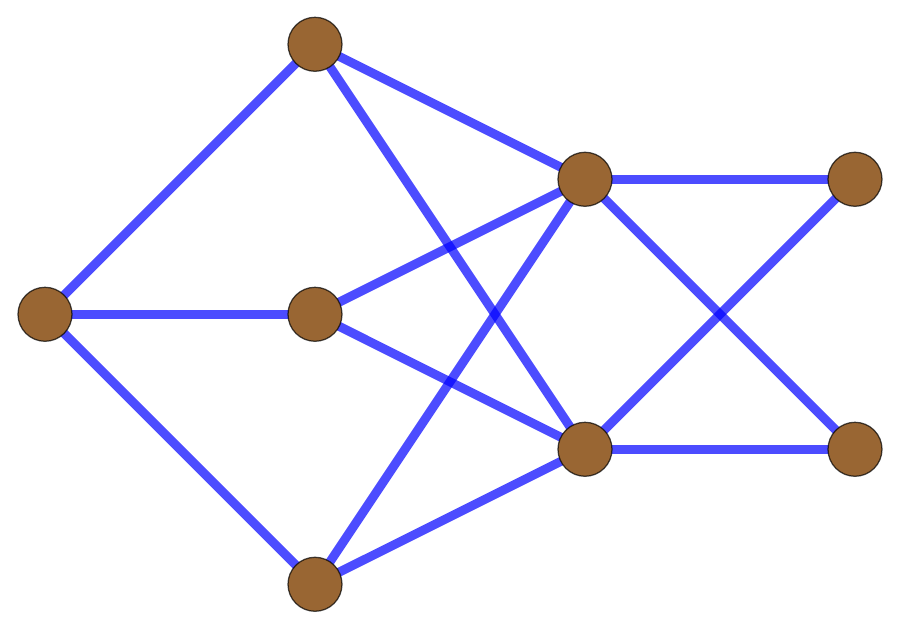}} \qquad
   $1241$: \scalebox{0.4}[0.4]{\includegraphics{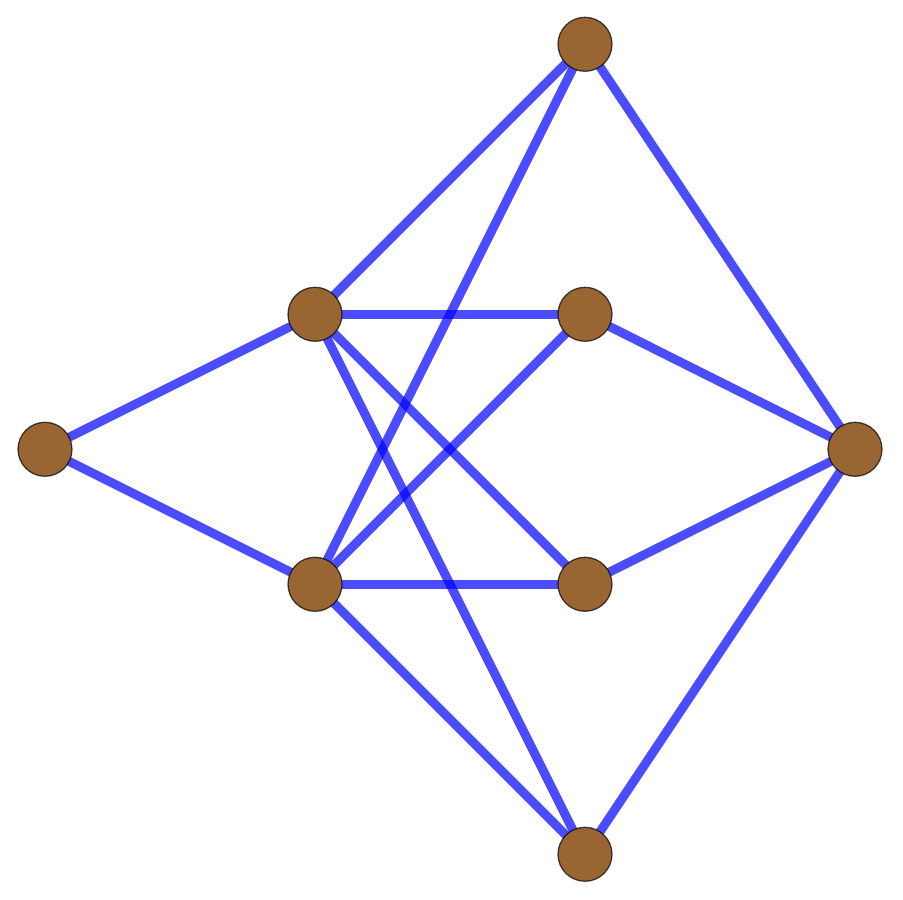}} \\
     \hline
     $2222$: \scalebox{0.4}[0.4]{\includegraphics{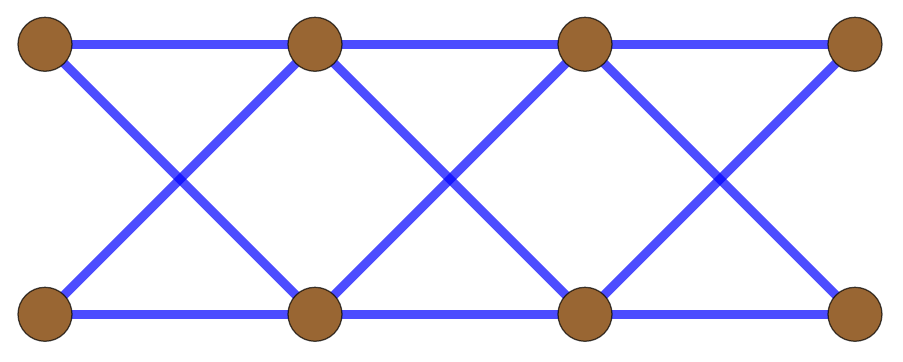}}  \qquad
      $12221$: \scalebox{0.5}[0.5]{\includegraphics{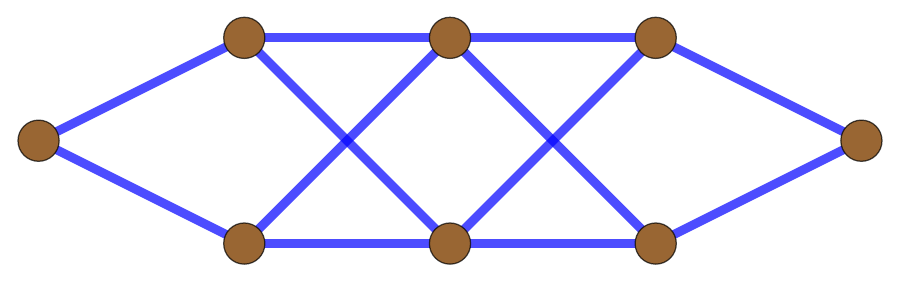}}\\
  \hline
  \hline
\end{tabular}
\end{table}

In Tables 1 and 2, we list all the graphs with $N\le 8$ vertices which are connected and satisfy the ``length-$2$ path'' and the ``no $3$-cycle'' properties stated near the end of Section IIB. We make the requirement of connectedness since disconnected graphs correspond to models with decoupled sets of states, and they can be constructed trivially from connected graphs. We can see that the number of such graphs increases rapidly with the number of vertices $N$. Several of these graphs correspond to MTLZ models that are previously identified, and some other graphs have been checked in \cite{MTLZ} to have no solutions. The $K_{1,1}$ graph (in the first row of Table 1) is the well-known two-state LZ model that was solved in the 1930s \cite{landau,zener,majorana,stuckelberg}. Other solvable graphs are named ``square'', ``cube'', and ``fans''. The $K_{2,2}$ graph (in the second row of Table 1) corresponds to a solvable $4$-state model which was first discovered in \cite{4-state-2002}; its another phase was later identified in \cite{4-state-2015}. In \cite{MTLZ} it was called the ``square'' model. The $8$-state ``cube'' graph (the last graph in the first row of Table 2) was considered in detail in \cite{MTLZ}. The square and cube graphs are special cases of the ``hypercube'' graph \cite{MTLZ}, a graph that can be constructed by a direct product of a number $D\ge 2$ of two-state LZ models (The two-state LZ model itself can also be viewed as the $D=1$ section of the hypercubes). The $K_{2,n}$ ($n=3,\ldots,6$) graphs (each first graph of the $5\le N\le 8$ graphs in Tables 1 and 2) belong to the ``fan'' model discovered in \cite{large-class}. Except the hypercubes, the fans and models constructed by their direct products, other MTLZ models have not been found. It was conjectured in \cite{MTLZ} that there are just no more such MTLZ models, but it is difficult to prove this given the large number of possible graphs.

We now apply the no-go rules on the graphs in Tables 1 and 2 with $ N\le 8$ vertices. 
It's easy to see that the graphs $K_{3,3}$, $K_{3,4}$, $1231$, $K_{3,5}$, $K_{4,4}$, cube$+2$, cube$+3$, $1232$, $1232$$-1$, $1322$, $1241$ are forbidden by the ``no $K_{3,3}$'' rule, and the graphs $1221$, $1222$, $1223$, $2222$, $12221$ are forbidden by the ``no $1221$'' rule. The graphs $K_{2,2}$, $K_{2,n}$ ($n=3,\ldots,6$) and cube are allowed by the two no-go rules, and they indeed support MTLZ models as described before. There is only one exception: the cube$+1$ graph is also allowed by the two rules, but detailed analysis shows that it does not support a solution. 
Thus, we conclude that for graphs with no more than $8$ states, the square, the cube and the fans indeed are the only solvable graphs.

The above application of the no-go rules on graphs with $N\le 8$ shows that these rules are powerful in identifying graphs which have no solutions. In fact, some graphs in Tables 1 and 2 were considered in \cite{MTLZ}. The $1221$, $K_{3,3}$, $2222$ graphs, named ``square with ears'', ``Mobius ladder'' and ``double-fan'' respectively in \cite{MTLZ}, were proven to have no solutions. The cube$+2$ and cube$+3$ graphs were also considered in \cite{MTLZ}; no solutions were found but there wasn't a proof that these two graphs do not support solutions. Now by the no-go rules, we can quickly judge that all these graphs are not solvable.

We expect that the no-go rules are also powerful in searching of new MTLZ models since they help ruling out a large number of graphs with no solutions. 
In the next subsection we are going to discuss this point.

\subsection{A scheme of graph classification}
For larger graphs with $N\ge 9$ vertices, are there new MTLZ models? According to the above analysis on  $N\le 8$ graphs, although many graphs are forbidden by the no-go rules, there can still be graphs like the cube$+1$ graph which are allowed by the no-go rules and does not belong to the hypercubes, the fans or their direct products. Whether such graphs can support solvable models require further investigation. Here, we are going to describe a scheme of classification of possible graphs at any $N$ which allows convenient usage of the no-go rules. This scheme can serve as a guideline of the search of MTLZ models.

\begin{figure}[!htb]
\scalebox{0.7}[0.7]{\includegraphics{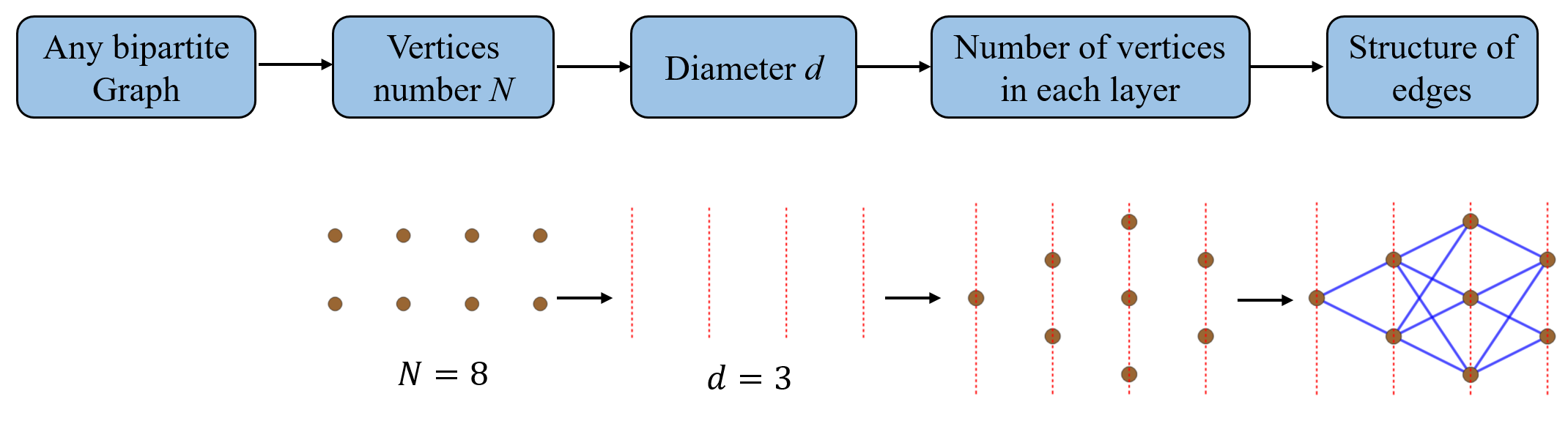}}
\caption{Illustration of the scheme of graph classification for MTLZ models. The upper panel is a flowchart of this scheme, and the lower panel is an example of identification of a specific graph by applying this scheme.}
\label{fig:scheme}
\end{figure}

The scheme is based on the ``layer graph'' notation introduced at the beginning of Section III. Let's consider graphs with a fixed vertex number $N$. We first classify them in terms of their {\it diameters} $d$ \cite{note-diameter}. A graph with a given diameter $d$ can be drawn as a layer graph with $d+1$ layers -- we call them the $0$th layer, the $1$st layer to the $d$th layer. We next fix the numbers of vertices in each layer, which we denote as $N_0, N_1,\ldots, N_d$ for the $0$th, the $1$st up to the $d$th layer, with $N_0+N_1+ \ldots+ N_d=N$. We then consider all possibilities of edges connecting adjacent layers. The distribution of vertices and edges should be allowed by the no-go rules and should also satisfy the ``length-$2$ path'' property. Such a scheme is illustrated in Fig.~\ref{fig:scheme}.

The above scheme needs justification in two places. First, does it enumerate all possible graphs? This amounts to the question that if any graph can be drawn as a layer graph. The answer is actually no -- a graph can be drawn as a layer graph if and only if it is {\it bipartite}. We can see this by simple arguments: a layer graph is bipartite since we can always put all the layers with even indices in a group and all the layers with odd indices in another group; conversely, a bipartite graph can always be drawn as a layer graph with two layers, each layer corresponding to each of the two groups. So the scheme enumerates all bipartite graphs but does not take into account any non-bipartite graphs \cite{note-odd cycles}.
It turns out that for graphs with vertices not more than $10$, this is not an issue -- there is simply no non-bipartite graphs with $N\le 10$ which are solvable. A theorem in the graph theory states that a graph is bipartite if and only if it does not contain any odd cycle, namely, a cycle of odd length \cite{Diestel}. For graphs for MTLZ models, by the ``no $3$-cycle'' property $3$-cycles cannot exist, so the minimal length of an odd cycle is $5$. Starting from a $5$-cycle, one finds that to satisfy the ``length-$2$ path'' property more vertices must be added, and the minimal model is a ``double-pentagon'' graph with $N=10$ vertices. This graph was considered in \cite{MTLZ} and proved to be not solvable. Here we can easily arrive at this proof, since this graph is forbidden by the no $1221$ rule. (The same goes for a general ``double-polygon'' graph with cycles of lengths longer than $5$.) It is not difficult to see that any other non-bipartite graph must contain more than $10$ vertices, so no $N\le 10$ non-bipartite graphs are solvable.

Second, can any bipartite graph with diameter $d$ be drawn as a layer graph with $d+1$ layers? The answer is positive since we can construct such a layer graph in a definite way. We first choose a length-$d$ path in the graph and put one of its ending vertex (let's call it $v$) in the $0$th layer. Then we put any other vertex $a$ with distance $d_{v,a}$ from the vertex $v$ to be in the $d_{v,a}$th layer. To prove that the graph constructed in this way is a layer graph, we need to show that any two vertices not in adjacent layers are not connected by an edge, namely, any two different vertices $a$ and $b$ in the $d_{v,a}$th and $d_{v,b}$th layers do not have an edge between them if $|d_{v,a}-d_{v,b}| \ge 2$ or if $d_{v,a}=d_{v,b}$. By the construction, their distances from vertex $v$ are $d_{v,a}$ and $d_{v,b}$, respectively. Assume without loss of generality that $d_{v,a}\ge d_{v,b}$. If $d_{v,a}-d_{v,b} \ge 2$, then if $a$ and $b$ are connected by an edge, there will be a path from $v$ through $b$ to $a$ with length $d_{v,b}+1$. Since $d_{v,a}-d_{v,b} \ge 2$, this length is a number not larger than $d_{v,a}-1$, so this contradicts the assumption that the distance between the vertices $a$ and $v$ is $d_{v,a}$. If $d_{v,a}=d_{v,b}$, then if $a$ and $b$ are connected by an edge, there will be a cycle from $v$ through $a$ through $b$ to $v$, which length is $d_{v,a}+d_{v,b}+1=2d_{v,a}+1$, an odd number. This contradicts the assumption that the considered graph is bipartite, since any graph which has an odd cycle is non-bipartite \cite{Diestel}. Therefore, any two vertices $a$ and $b$ with $|d_{v,a}-d_{v,b}| \ge 2$ or $d_{v,a}=d_{v,b}$ must not have an edge between them, and the constructed graph is a layer graph. Note that when applying such a construction procedure the ways of representation of a graph as a layer graph may not be unique, since a graph with diameter $d$ may contain more than one length-$d$ paths. This leads to extra work when enumerating possible graphs, but it is not a big issue -- it is more important that the procedure indeed enumerates all the possibilities, and no graphs will be missed.

Such a ``layer graph'' scheme turns out to be convenient for application of the no-go rules and also the ``length-$2$ path'' and ``no $3$-cycle'' properties. The ``no $3$-cycle'' property is automatically satisfied due to our definition of a layer graph. The ``length-$2$ path'' property requires that any ``inner'' layers (any layer except the $0$th or the $d$th layer) must contain at least $2$ vertices, namely,  $N_1, \ldots ,N_{d-1}\ge 2$. The no $1221$ rule requires that two adjacent inner layers must not both have $2$ vertices, namely, the sequence $N_0  N_1 \ldots  N_d$ must not contain ``$\ldots 22 \ldots$'' parts. The no $K_{3,3}$ rule further requires that a layer with at least $3$ vertices must not be too connected to its adjacent layers. Given any layer graph, one can quickly judge whether the above requirements are satisfied, and thus whether the graph is possible to host MTLZ models. Furthermore, the scheme is systematic and simple enough and we expect that it can be conveniently automated by computer programming.

\begin{figure}[!htb]
(a)~ \scalebox{0.5}[0.5]{\includegraphics{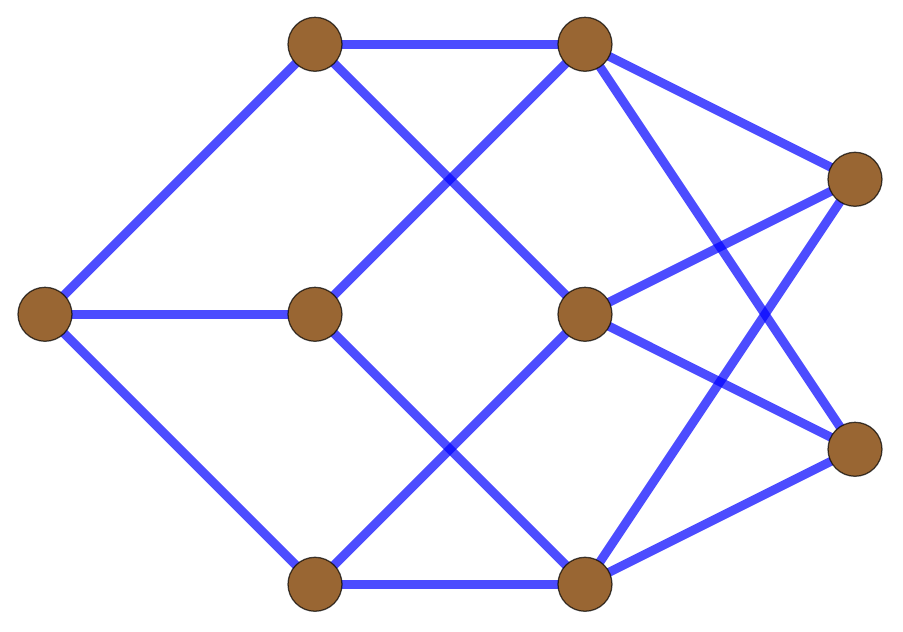}}  ~ ~ ~ ~ ~ ~ ~ ~ ~ ~ ~
(b)~ \scalebox{0.5}[0.5]{\includegraphics{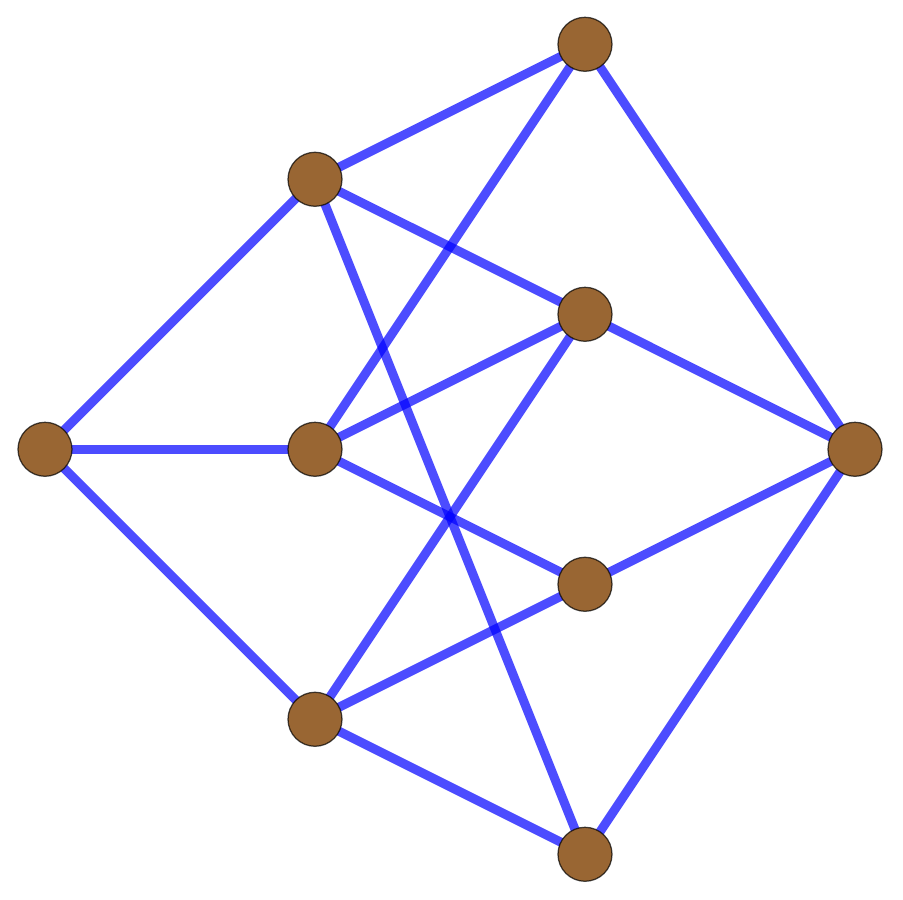}}
\caption{All possible connected graphs with $N=9$ vertices that satisfy the ``length-$2$ path'' property, the ``no $3$-cycle'' property and are allowed by the no-go rules.}
\label{fig:N9}
\end{figure}

Applying the above classification scheme, we've drawn out all possible graphs for $N=9$ and $N=10$. 
For $N=9$ we find only two graphs that are allowed by the no-go rules, as shown in Fig.~\ref{fig:N9}. However, detailed analysis, which we present in the appendix, shows that these two graph do not support solutions either. Therefore, we arrive at the conclusion that for MTLZ models with no more than $9$ states, besides the hypercubes and the fans, there are no solvable graphs. For $N=10$ the number of allowed graph is much larger -- we find $15$ graphs. Whether these graphs may support solutions requires future study.

\section{Conclusions and Perspectives}
We studied a class of exactly solvable quantum many-body models, named the multitime Landau-Zener (MTLZ) model, which is multitstate and multitime generalization of the two-state Landau-Zener model. Parameters of an MTLZ model can be represented as certain types of data on a graph. 
We proved two no-go rules which strongly restrict structures of graphs that could host MTLZ models. We then applied the no-go rules to show that for models with no more than $9$ states, besides the hypercubes and the fans discovered previously, there are no other MTLZ models. We also proposed a scheme to systematically classify graphs that could possibly support MTLZ models, which could serve as a guideline to look for new MTLZ models. 

It remains an open question whether there are new MTLZ models with $N> 9$ states; future work could start with analysis on models with $N=10$ states. It is also interesting to look for new no-go rules for graphs of MTLZ models that are independent of the two no-go rules presented here. It's possible that they do exist, but the structures of the involved graphs  may be more complicated and applications of them may not be as easy as the two rules presented here. Studies along this line may lead to discoveries of new solvable models or, in the opposite direction, an ultimate proof of the conjecture that the solvable graphs currently found (the hypercubes, the fans and graphs corresponding to direct products of these models) are {\it the} only graphs that support MTLZ models.

\section*{Acknowledgements}
The authors are grateful for discussions with Nikolai A.~Sinitsyn. This work was supported by NSFC (No. 12105094) and by the Fundamental Research Funds for the Central Universities from China.

\section*{Conflict of Interest and Data Availability Statements}
On behalf of all authors, the corresponding author states that there is no conflict of interest, and that this manuscript has no associated data. 

\newpage
\section*{Appendix: Proof of no solutions in two graphs with $N=9$}

\setcounter{figure}{0}
\setcounter{equation}{0}
\renewcommand{\theequation}{A\arabic{equation}}
\renewcommand\thefigure{A\arabic{figure}}

In this appendix, we prove that two certain graphs with $N=9$ which are allowed by the no-go rules are still not solvable, namely, they do not support MTLZ models.

\subsection{Proof for the graph in Fig.~\ref{fig:N9}(a)}
\begin{figure}[!htb]
	\scalebox{0.6}[0.6]{\includegraphics{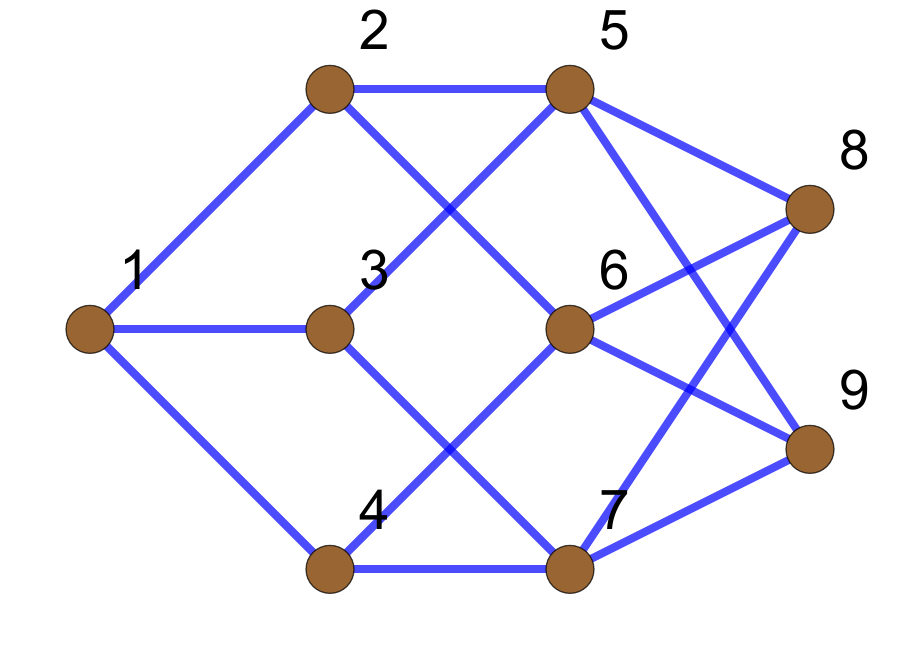}}
	\caption{The graph in Fig.~\ref{fig:N9}(a) with vertices labelled by numbers from $1$ to $9$.}
	\label{fig:1332}
\end{figure}

FIG.~\ref{fig:1332} is the same as Fig.~\ref{fig:N9}(a), but with vertices labelled by numbers. We first consider the $4$-cycle formed by the vertices $2$, $5$, $6$ and $8$, i.e. the cycle $2568$. 
The wedge product relations of $\bar{A}^{ab}$ forms on its edges read:
\begin{eqnarray}
&\bar{A}^{25}\wedge \bar{A}^{58}=r_{2586}\bar{A}^{26}\wedge \bar{A}^{68},\\
&\bar{A}^{25}\wedge \bar{A}^{26}=r_{5268}\bar{A}^{58}\wedge \bar{A}^{68}.
\end{eqnarray}
The length-$2$ path condition 
between vertices $2$ and $8$ further requires that $r_{2586}=-1$. For the cycle $2569$, due to the equivalence of vertices $8$ and $9$, we can get
\begin{eqnarray}
\bar{A}^{25}\wedge \bar{A}^{26}=r_{5269}\bar{A}^{59}\wedge \bar{A}^{69},\quad r_{2596}=-1.
\end{eqnarray}
For the cycle $5869$, we have
\begin{eqnarray}
&\bar{A}^{58}\wedge \bar{A}^{68}=r_{5869}\bar{A}^{59}\wedge \bar{A}^{69},\\
&\bar{A}^{58}\wedge \bar{A}^{59}=r_{8596}\bar{A}^{68}\wedge \bar{A}^{69}.
\end{eqnarray}
The above wedge product relations among $\bar{A}^{25}\wedge \bar{A}^{26}$,  $\bar{A}^{58}\wedge \bar{A}^{68}$  and  $\bar{A}^{59}\wedge \bar{A}^{69}$ imply that
\begin{eqnarray}
r_{5268}r_{5269}r_{5869}=1.
\label{r1}
\end{eqnarray}
Similarly, for the cycles $6478$, $6479$ and $6879$, the wedge product relations among $\bar{A}^{46}\wedge \bar{A}^{47}$,  $\bar{A}^{68}\wedge \bar{A}^{78}$  and  $\bar{A}^{69}\wedge \bar{A}^{79}$ suggest that
\begin{eqnarray}
r_{6478}r_{6479}r_{6879}=1.
\label{r2}
\end{eqnarray}
The length-$2$ path conditions between $4$ and $8$ and between $4$ and $9$ give $r_{4687}=r_{4697}=-1$. And for the cycles $5378$, $5379$ and $5879$, the wedge product relations among $\bar{A}^{35}\wedge \bar{A}^{37}$,  $\bar{A}^{58}\wedge \bar{A}^{78}$  and  $\bar{A}^{59}\wedge \bar{A}^{79}$ suggest that
 \begin{eqnarray}
 r_{5378}r_{5379}r_{5879}=1.
 \label{r4}
 \end{eqnarray}
The length-$2$ path conditions between $3$ and $8$ and between $3$ and $9$ give $r_{3587}=r_{3597}=-1$. Finally, for the cycles $5869$, $6879$ and $5879$, the wedge product relations among $\bar{A}^{58}\wedge \bar{A}^{59}$,  $\bar{A}^{68}\wedge \bar{A}^{69}$  and  $\bar{A}^{78}\wedge \bar{A}^{79}$ suggest that
 \begin{eqnarray}
 r_{8596}r_{8697}r_{8597}=1.
 \label{r3}
 \end{eqnarray}

Multiplying Eqs. \eqref{r1}, \eqref{r2}, \eqref{r4}, \eqref{r3} and the above identified signs $r_{2586}=r_{2596}=r_{4687}=r_{4697}=r_{3587}=r_{3597}=-1$, we get
\begin{eqnarray}
  r_{5268}r_{2586}r_{5269}r_{2596}r_{6478}r_{4687}r_{6479}r_{4697}r_{5378}r_{3587}r_{5379}r_{3597}r_{5869}r_{8596}r_{6879}r_{8697}r_{5879}r_{8597}=1.
  \label{r5}
 \end{eqnarray}
This relation involves 18 signs for the 9 cycles $2568$, $2596$, $4687$, $4697$, $3587$, $3597$, $5869$, $6879$ and $5879$.
It does not depend on specific orientations of the $4$-cycles in the graph. Recall that the two signs $r_{abcd}$ and $r_{badc}$ corresponding to a cycle $abcd$ have different relations according to the orientation of the cycle:
  \begin{align}
  &r_{abcd}r_{badc}=1, \quad \textrm{if the cycle $abcd$ is of the non-bipartite orientation,} \label{eq:rnb1}\\
  &r_{abcd}r_{badc}=-1, \quad \textrm{if the cycle $abcd$ is of the bipartite orientation.} \label{eq:rb1}
  \end{align}
Therefore,  Eq.~\eqref{r5} requires that the number of cycles in the bipartite orientation among the $9$ cycles is even.

  \begin{figure}[!htb]
  	(a)~ \scalebox{0.5}[0.5]{\includegraphics{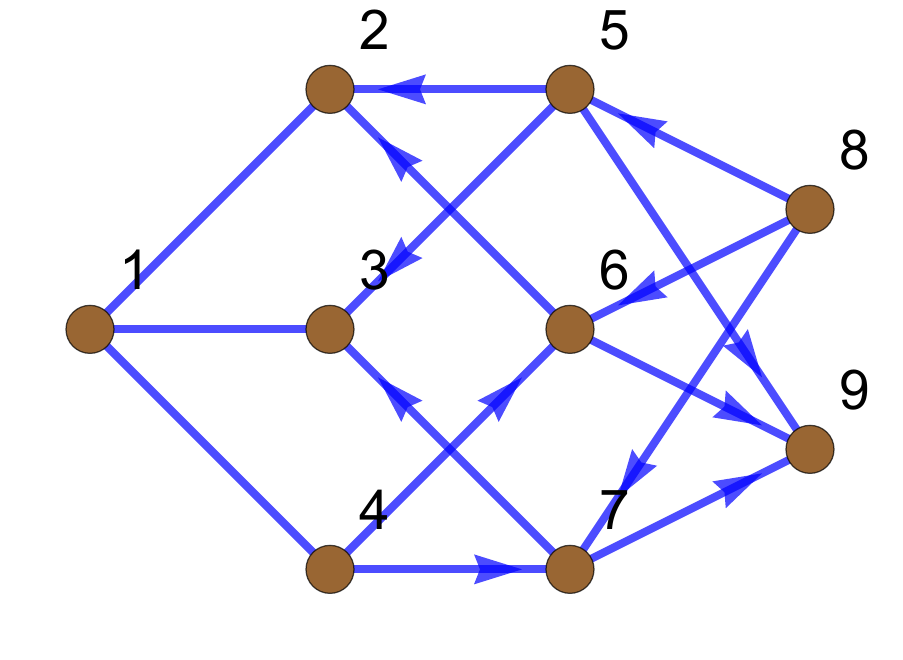}}
  	(b)~ \scalebox{0.5}[0.5]{\includegraphics{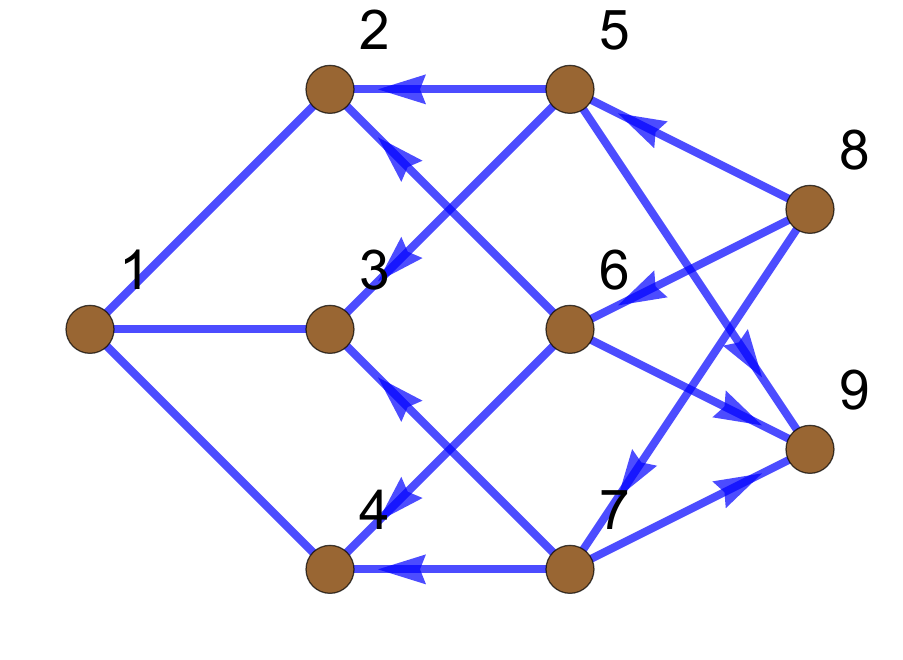}}
  	(c)~ \scalebox{0.5}[0.5]{\includegraphics{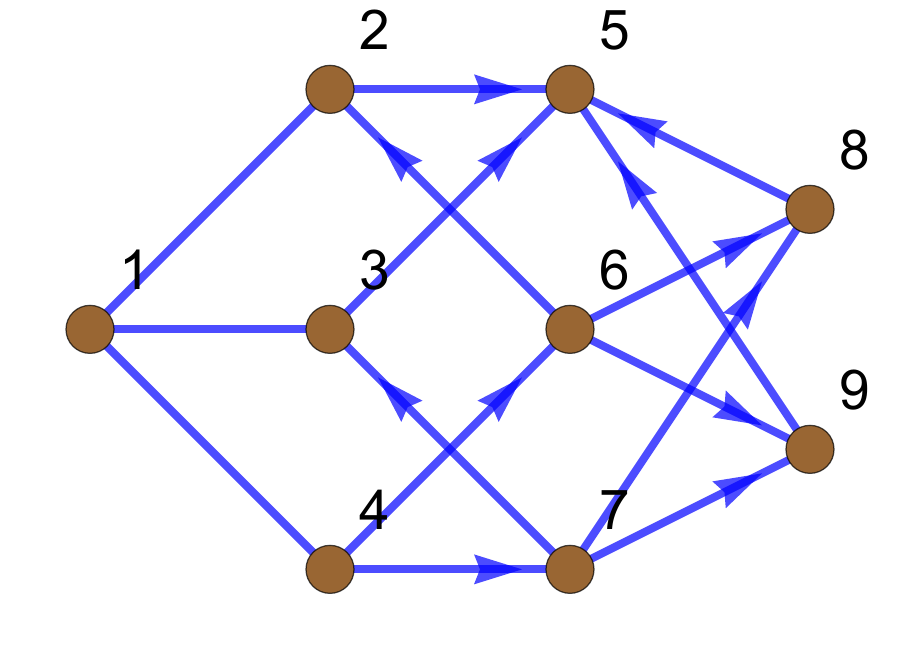}}
  	(d)~ \scalebox{0.5}[0.5]{\includegraphics{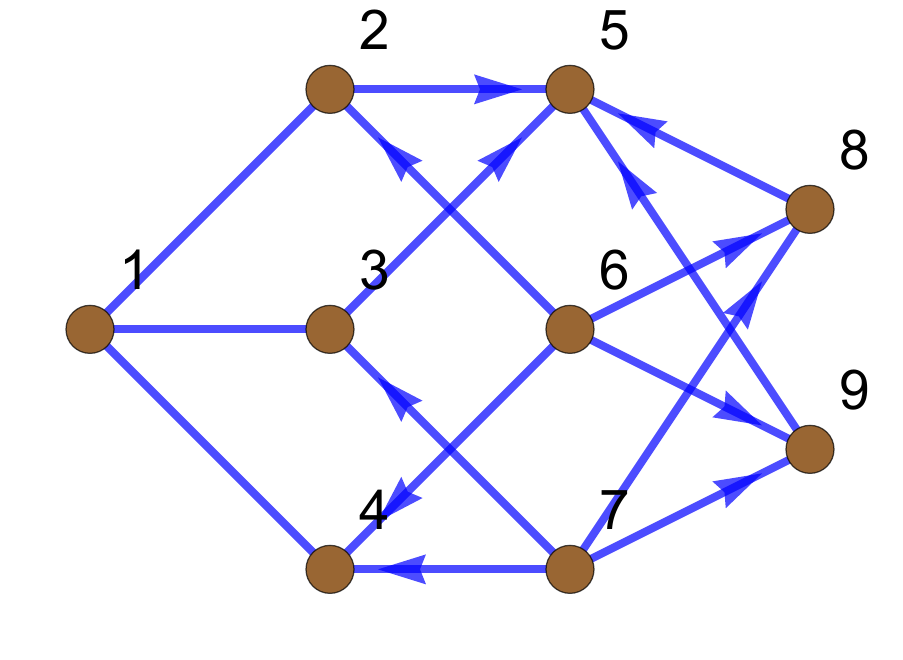}}
  	(e)~ \scalebox{0.5}[0.5]{\includegraphics{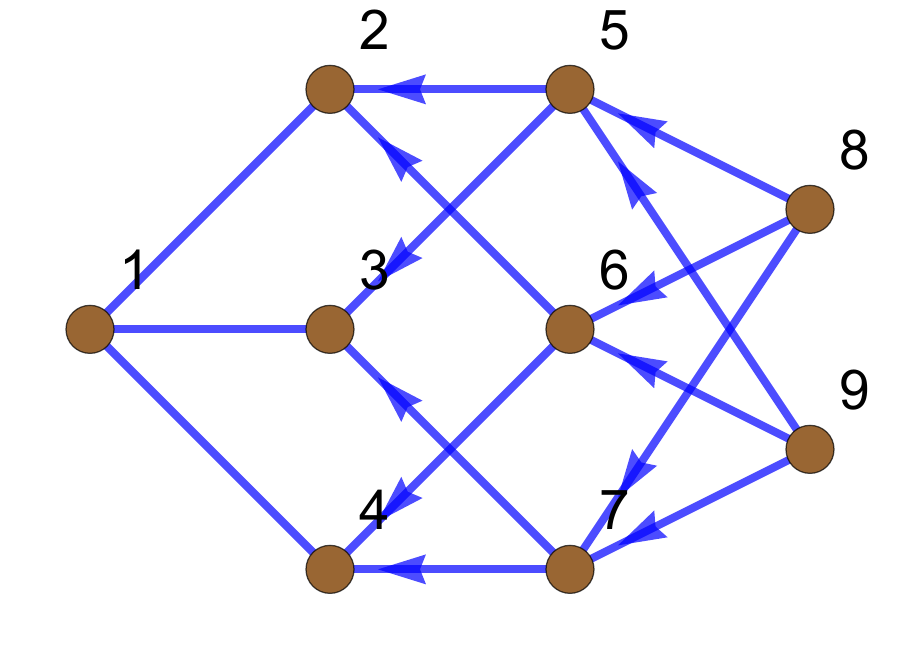}}
  	(f)~ \scalebox{0.5}[0.5]{\includegraphics{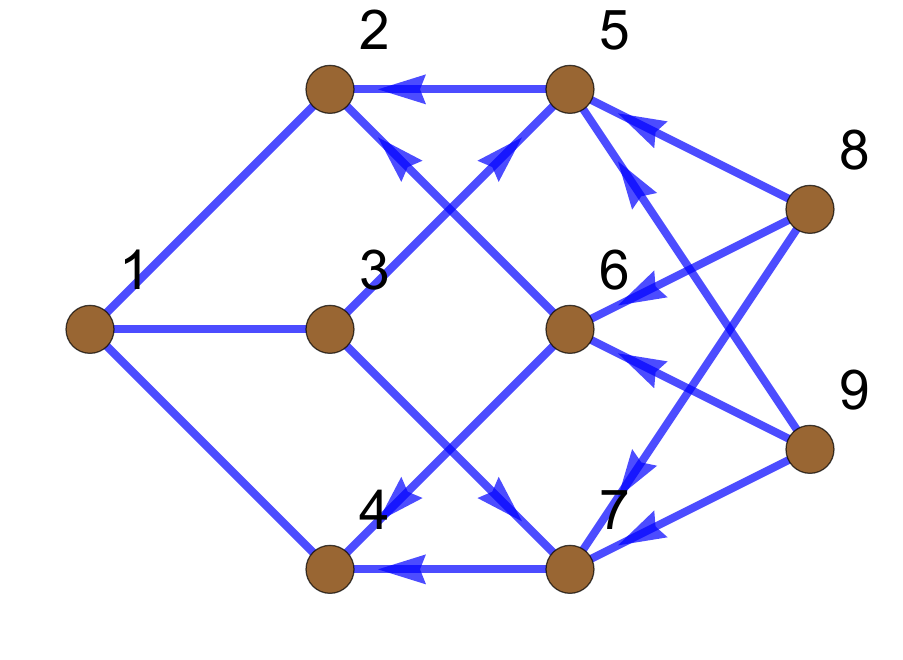}}
  	(g)~ \scalebox{0.5}[0.5]{\includegraphics{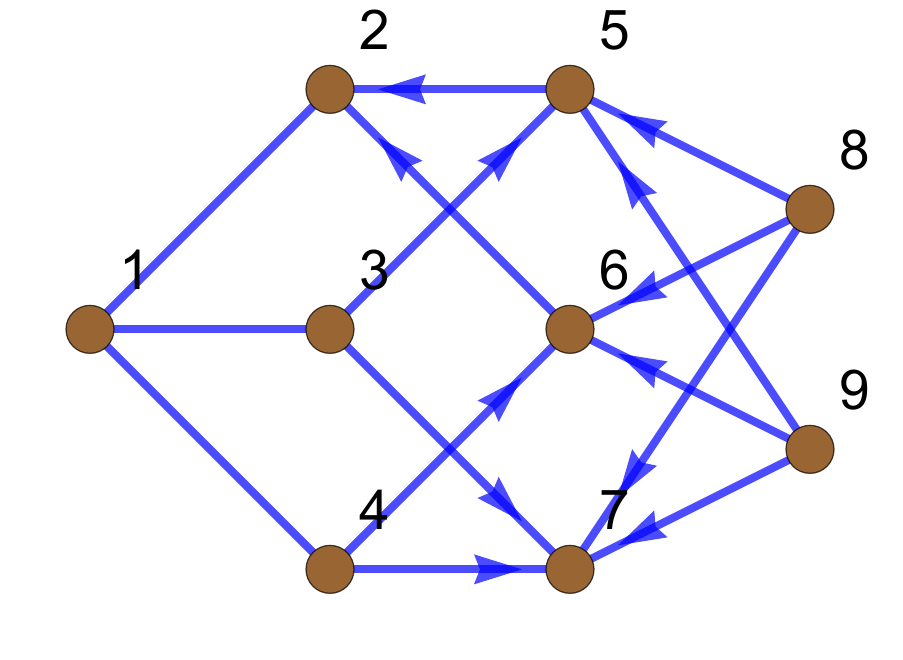}}
  	(h)~ \scalebox{0.5}[0.5]{\includegraphics{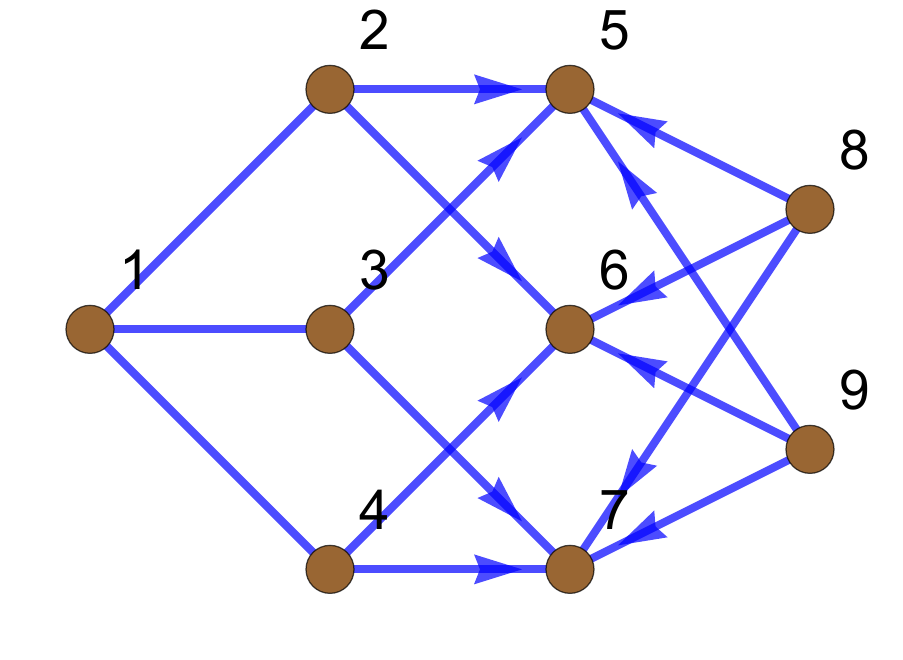}}
  	\caption{All possible cases of orientations for the cycles $2568$, $2596$, $4687$, $4697$, $3587$, $3597$, $5869$, $6879$, $5879$.}
  	\label{fig:1332-8}
  \end{figure}

We now consider specific orientations of these 9 cycles. It turns out that up to symmetry of vertices there are $8$ possible cases, as shown in FIG.~\ref{fig:1332-8}. By direct counting we see that in all these cases the numbers of cycles in the bipartite orientation among these $9$ cycles are odd. This means that for all cases we have
 \begin{eqnarray}
 r_{5268}r_{2586}r_{5269}r_{2596}r_{6478}r_{4687}r_{6479}r_{4697}r_{5378}r_{3587}r_{5379}r_{3597}r_{5869}r_{8596}r_{6879}r_{8697}r_{5879}r_{8597}=-1.
  \label{r6}
 \end{eqnarray}
Eqs.~\eqref{r5} and \eqref{r6} are contradictory, so the graph in Fig.~\ref{fig:1332} is not solvable.

\subsection{Proof for the graph in Fig.~\ref{fig:N9}(b)}

\begin{figure}[!htb]
	(a)~ \scalebox{0.5}[0.5]{\includegraphics{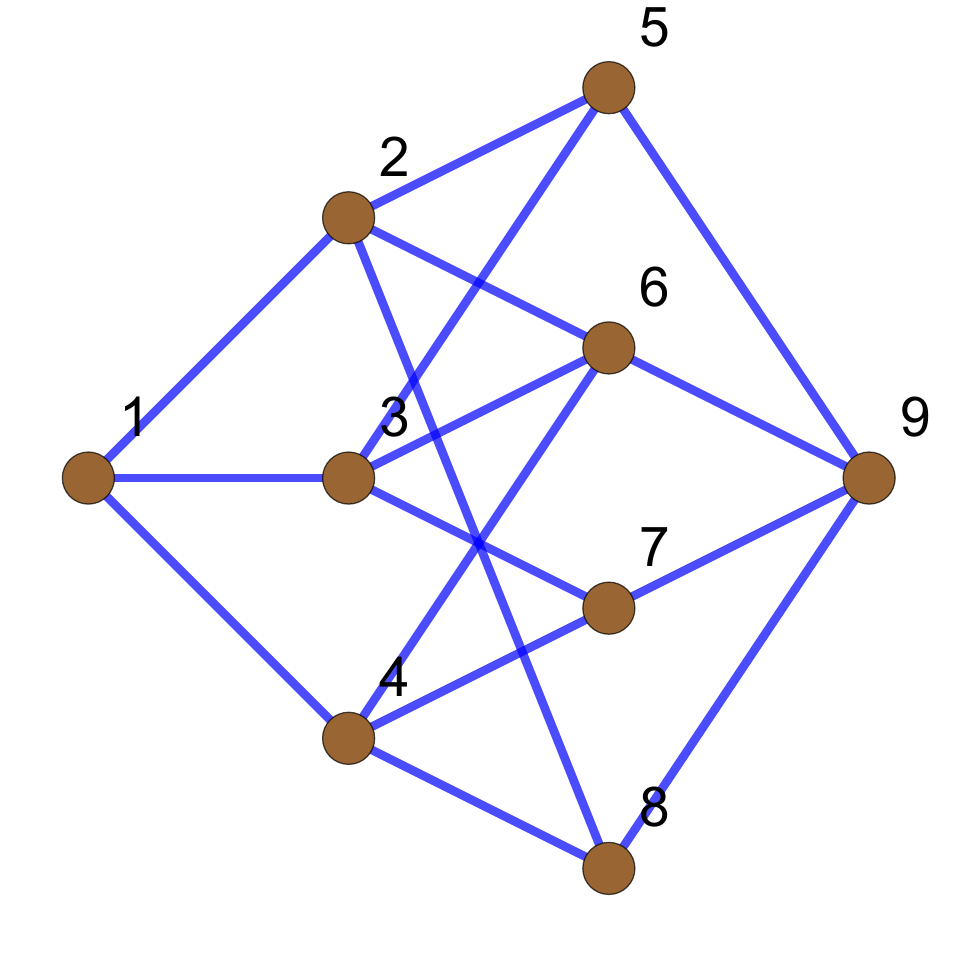}}  ~ ~ ~ ~ ~ ~ ~ ~ ~ ~ ~
	(b)~ \scalebox{0.5}[0.5]{\includegraphics{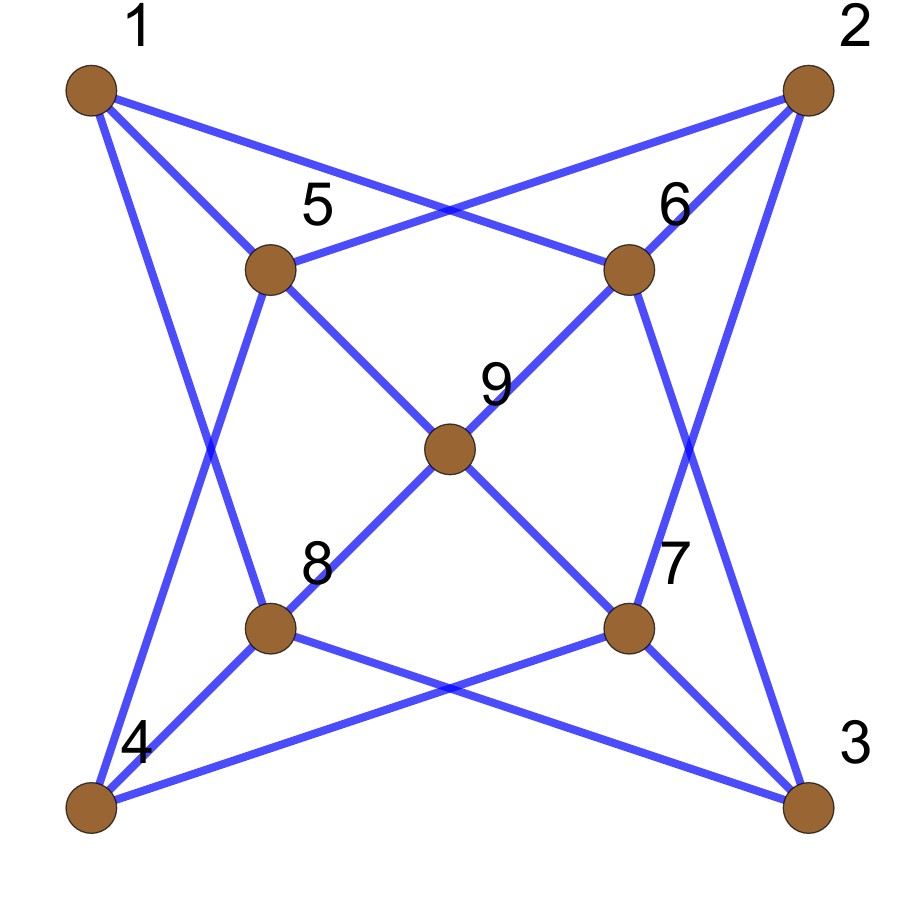}}
	\caption{(a) The graph in Fig.~\ref{fig:N9}(b) with vertices labelled by numbers from $1$ to $9$. (b) A graph equivalent to (a) which illustrates the symmetry of the vertices. Note that the labels have been renumbered and do not correspond to those in (a).}
	\label{fig:1341 and 4-point star}
\end{figure}

FIG.~\ref{fig:1341 and 4-point star}(a) is the same as Fig.~\ref{fig:N9}(b), but with vertices labelled by numbers. It can be drawn as FIG.~\ref{fig:1341 and 4-point star}(b) (with vertices renumbered) which illustrates the high symmetry of the vertices. We first look at the fan subgraphs with 5 vertices in this graph -- there are $10$ such subgraphs. For the fan made by the vertices $1$, $6$, $9$, $8$ and $3$, the wedge product relations among $\bar{A}^{16}\wedge \bar{A}^{18}$, $\bar{A}^{36}\wedge \bar{A}^{38}$  and  $\bar{A}^{69}\wedge \bar{A}^{89}$ gives
\begin{eqnarray}
r_{6189}r_{6389}r_{6183}=1.
\label{r7}
\end{eqnarray}
The multipath property \eqref{constr-3-LA-bar-A} between vertices $6$ and $8$ reads:
\begin{align}
&\sqrt{|\gamma ^{16}\gamma ^{18}|}\bar{A}^{16}\wedge \bar{A}^{18}+\sqrt{|\gamma ^{69}\gamma ^{89}|}\bar{A}^{69}\wedge \bar{A}^{89}+\sqrt{|\gamma ^{36}\gamma ^{38}|}\bar{A}^{36}\wedge \bar{A}^{38}
=0,\label{r8}
\end{align}
which means that $r_{6189}$, $r_{6389}$ and $r_{6183}$ cannot be all $1$. In combination with Eq.~\eqref{r7}, we see that two of the three sign factors $r_{6189}$, $r_{6389}$ and $r_{6183}$ should be $-1$, and one of them is $1$. For all other fan subgraphs with $5$ vertices, similar argument gives
 \begin{eqnarray}
 &r_{1596}r_{1598}r_{1698}=1, \label{r9}\\
&r_{3697}r_{3897}r_{3698}=1, \label{r11}\\
  &r_{5279}r_{5479}r_{5274}=1,  \label{r12-1}\\
  & r_{2596}r_{2697}r_{2597}=1, \label{r12-2}\\
  & r_{4598}r_{4798}r_{4597}=1, \label{r12-3}\\
& r_{5169}r_{5269}r_{5162}=1,   \label{r13}\\
  &r_{6279}r_{6379}r_{6273}=1,   \label{r14-1}\\
  & r_{8379}r_{7489}r_{7384}=1,  \label{r14-2}\\
  & r_{5189}r_{5489}r_{5184}=1,   \label{r14-3}
  \end{eqnarray}
where for each equation  two of the three sign factors should be $-1$, and one of them $1$. Next, the multipath property applied between any pair of vertices among $1$, $2$, $3$ and $4$ give:
  \begin{eqnarray}
   r_{1638}= r_{2547}= r_{1526}= r_{2637}= r_{3748}= r_{1548}=-1.
  \label{r15}
  \end{eqnarray}

Combining Eqs. \eqref{r7} and  \eqref{r9} -- \eqref{r15}, we get:
  \begin{eqnarray}
  &\nonumber r_{1596}r_{5169}r_{1598}r_{5189}r_{3697}r_{6379}r_{3897}r_{8379}r_{2596}r_{5269}r_{2697}r_{6279}r_{4598}r_{5489}r_{4798}r_{7489}r_{1698}r_{6189}\\
  &r_{3698}r_{6389}r_{1638}r_{6183}r_{2597}r_{5279}r_{4597}r_{5479}r_{2547}r_{5274}r_{1526}r_{5162}r_{2637}r_{6273}r_{3748}r_{7384}r_{1548}r_{5184}=1.
  \label{r16}
  \end{eqnarray}
 This relation involves the $36$ signs for all the $4$-cycles in the graph -- there are $18$ such cycles. Eq.~\eqref{r16} requires that the number of cycles in the bipartite orientation among these cycles $18$ is even.

   \begin{figure}[!htb]
  	(a)~ \scalebox{0.4}[0.4]{\includegraphics{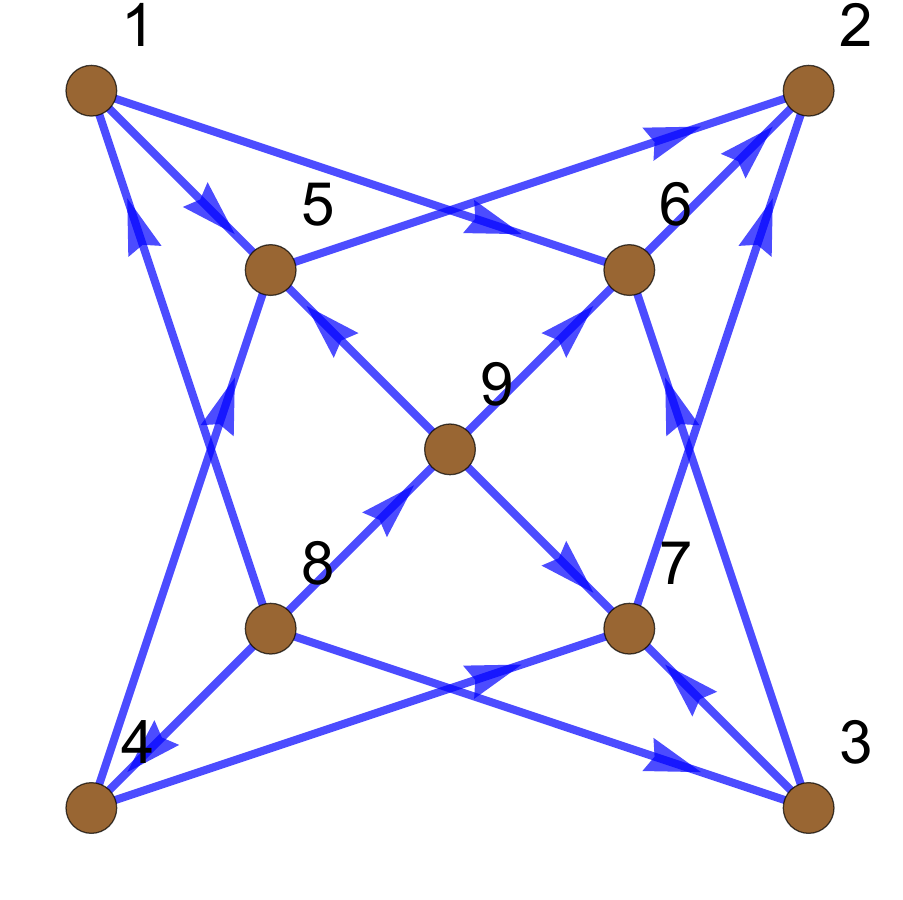}}
  	(b)~ \scalebox{0.4}[0.4]{\includegraphics{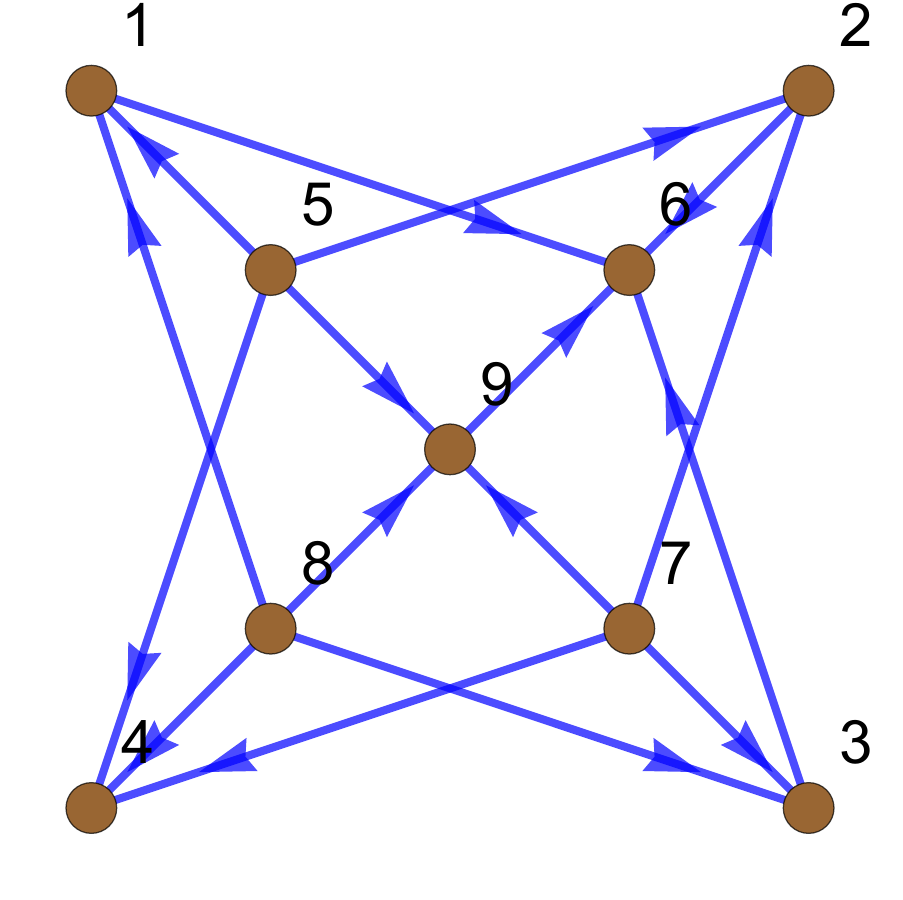}}
  	(c)~ \scalebox{0.4}[0.4]{\includegraphics{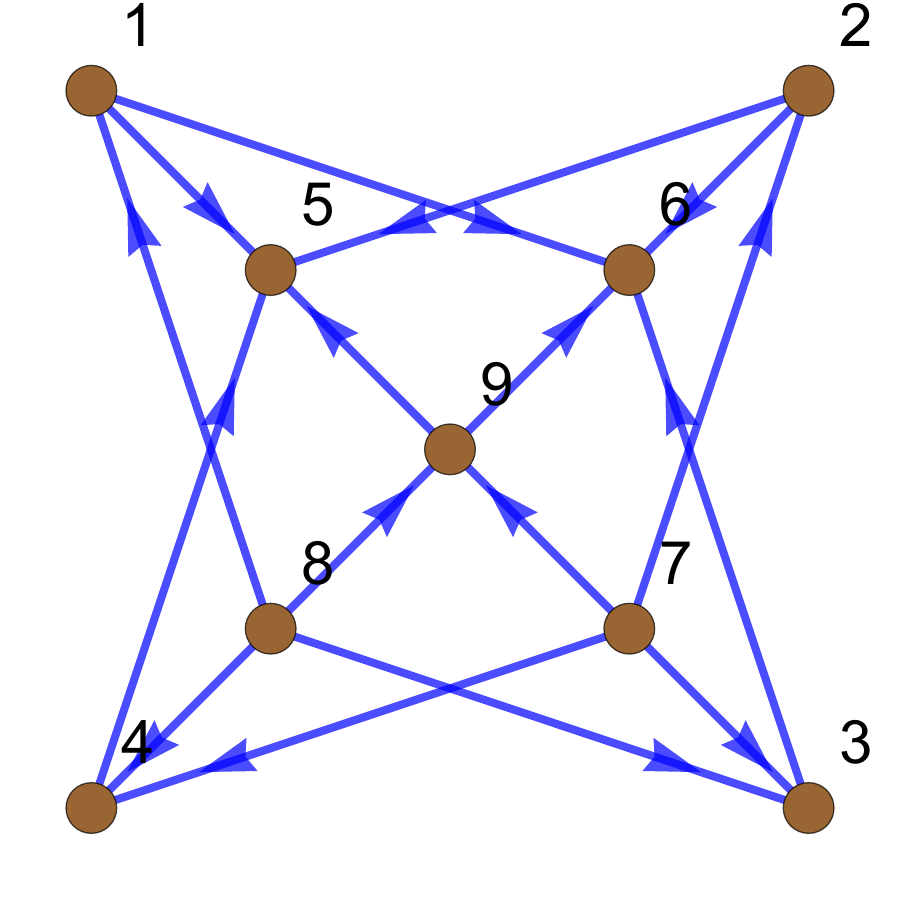}}
  	(d)~ \scalebox{0.4}[0.4]{\includegraphics{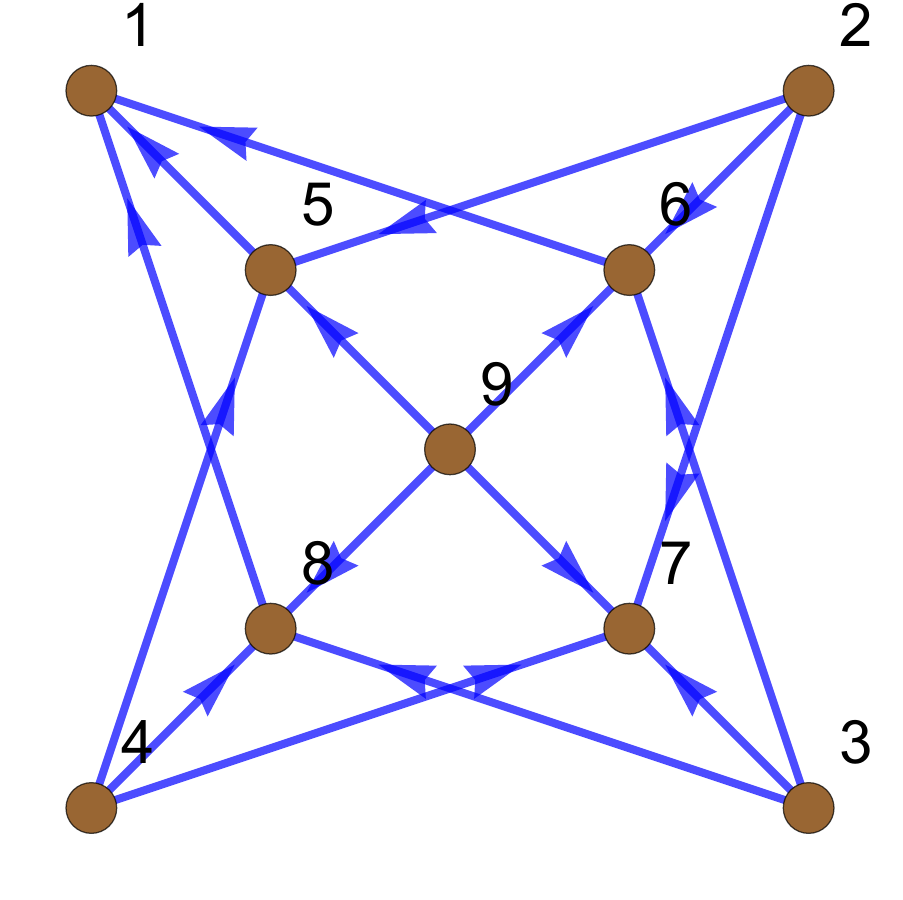}}
  	(e)~ \scalebox{0.4}[0.4]{\includegraphics{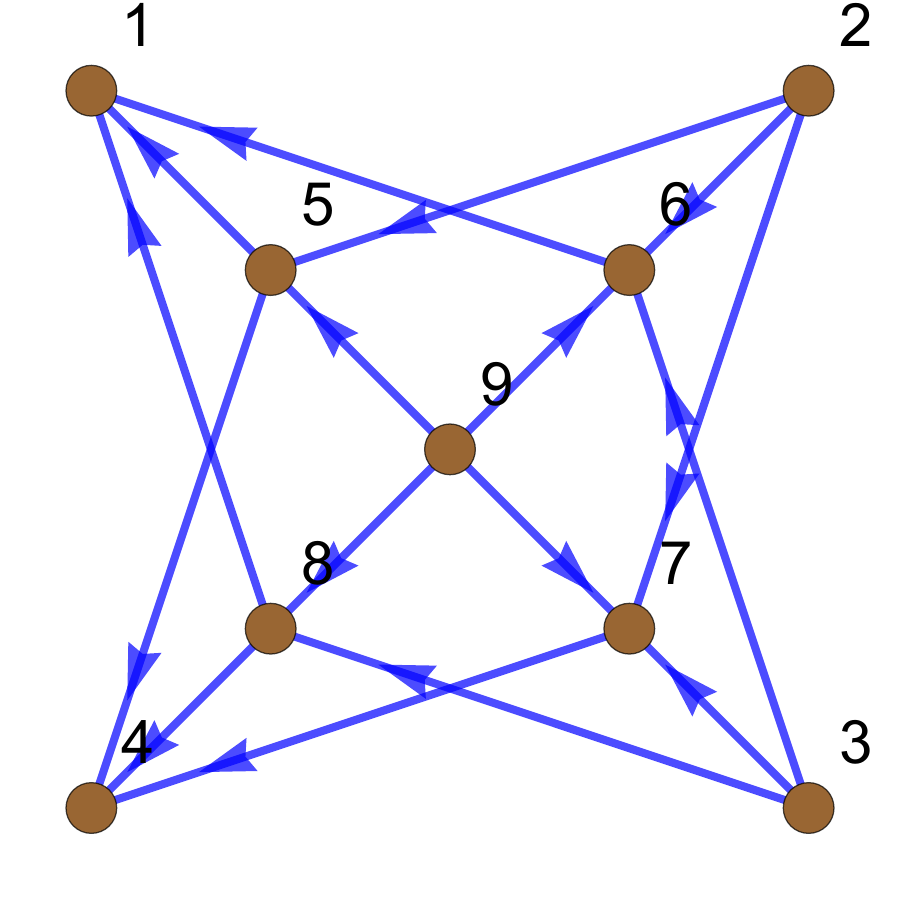}}
  	(f)~ \scalebox{0.4}[0.4]{\includegraphics{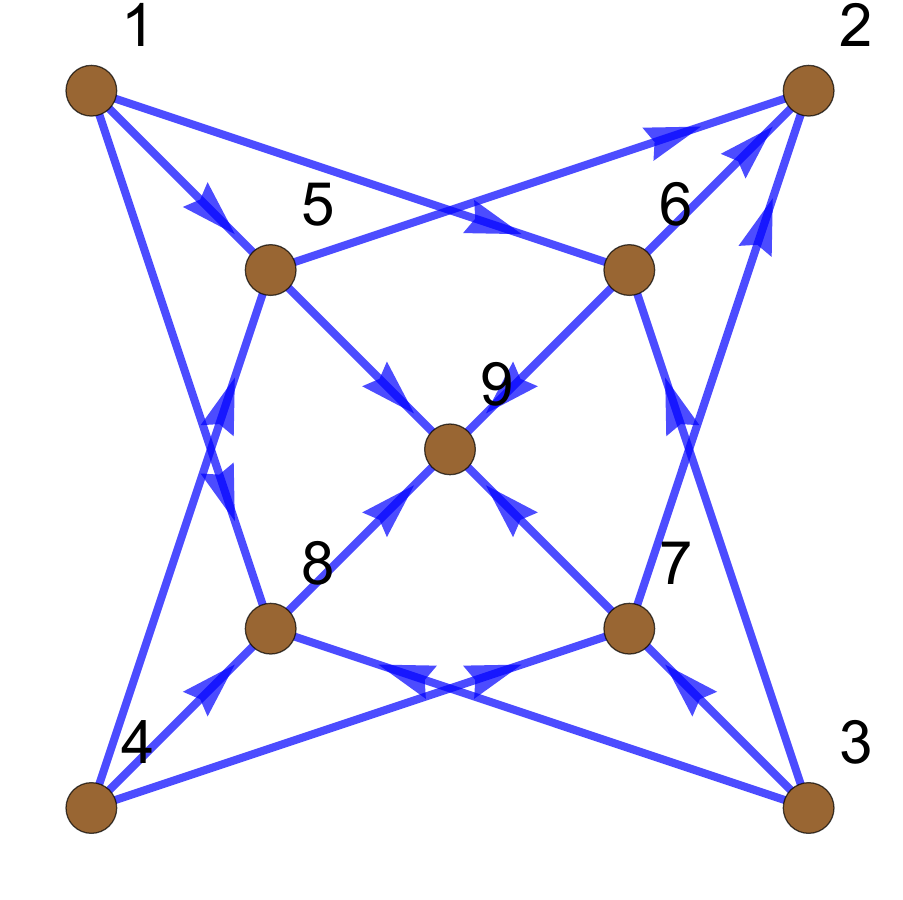}}
  	(g)~ \scalebox{0.4}[0.4]{\includegraphics{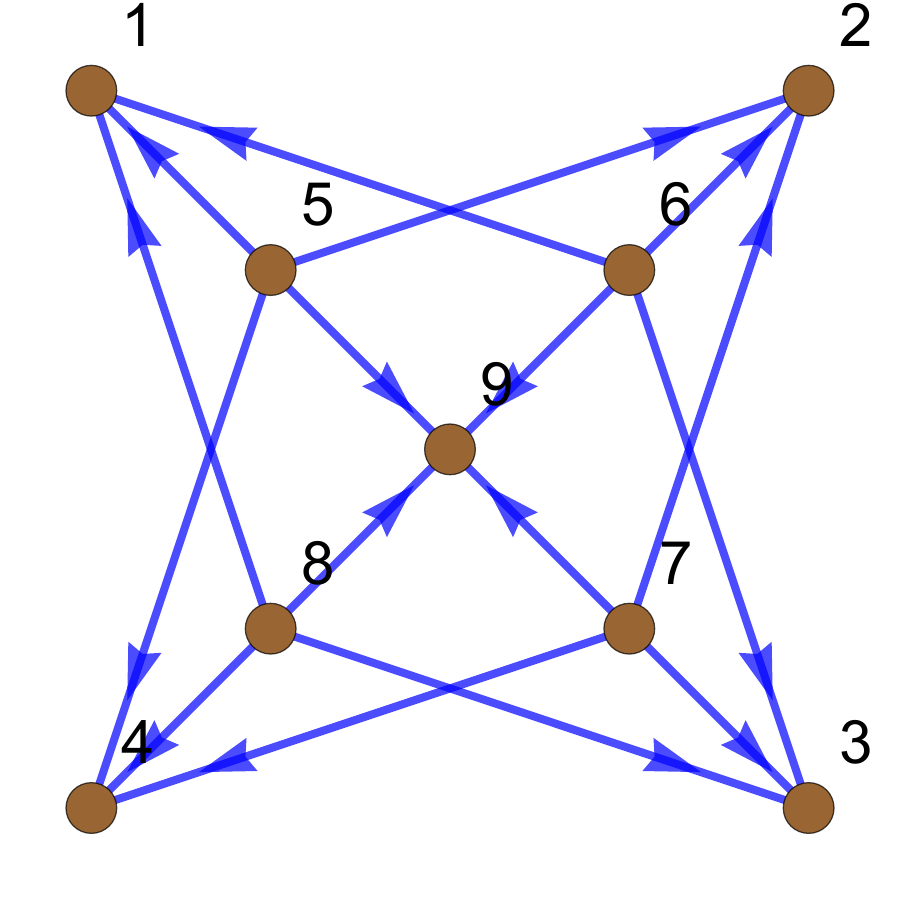}}
  	(h)~ \scalebox{0.4}[0.4]{\includegraphics{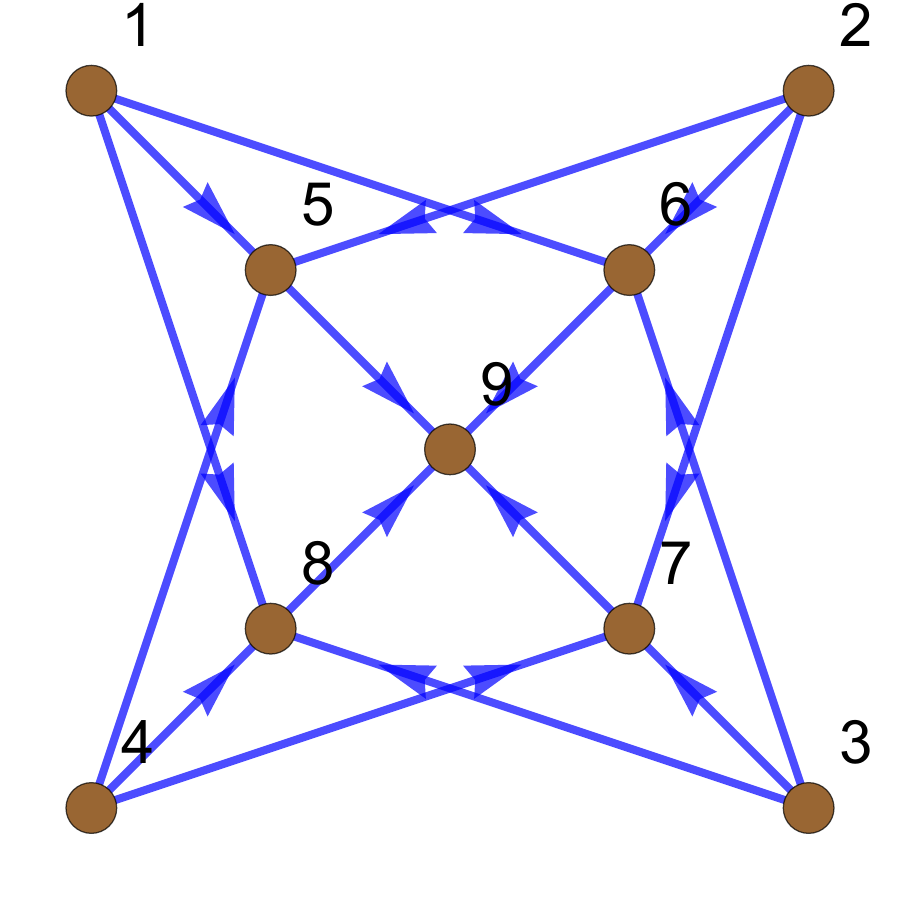}}
  	\caption{All possible cases of orientations for the graph in Fig.~\ref{fig:1341 and 4-point star}(b).}
  	\label{fig:4-point star}
  \end{figure}

We now consider specific orientations in this graph. There are $8$ cases, as shown in FIG.~\ref{fig:4-point star}. In cases (a) and (b), the numbers of cycles in the bipartite orientation among all the $18$ $4$-cycles are odd, so Eq.~\eqref{r16} is not satisfied and we can immediately conclude that these two cases do not support solutions. In the remaining $6$ cases Eq.~\eqref{r16} is satisfied, but detailed analysis still lead to contradictions. Once orientations are specified, we can determine a large number of sign factors through the relations mentioned above. For example, for case (c) the analysis goes as follows. The cycle $1526$ is bipartite, so from $r_{1526}=-1$ we get $r_{5162}=1$. Using Eq.~\eqref{r13}, we get $r_{5169}=r_{5269}=-1$. Since the cycle $5269$ is bipartite, we get $r_{2596}=1$. By  Eq.~\eqref{r12-2} we next get $r_{2697}=r_{2597}=-1$. Since the cycle $2697$ is non-bipartite, we have $r_{6279}=-1$. The cycle $2637$ is non-bipartite, so from $r_{2637}=-1$ we get $r_{6273}=-1$. Eq.~\eqref{r14-1} then requires that $r_{6379}=1$. The cycle $6379$ is non-bipartite, so $r_{3697}=1$, and we get $r_{3897}=r_{3698}=-1$ by Eq. \eqref{r11}. On the other hand, since the cycle $3748$ is bipartite, from $r_{3748}=-1$ we have $r_{3748}=1$, and then $r_{8379}=r_{7489}=-1$ by Eq. \eqref{r14-2}. And since the cycle $8379$ is bipartite, we get $r_{3897}=1$. This contradicts the previously obtained $r_{3897}=-1$, so case (c) does not support a solution. A similar argument for each of the rest cases (d) -- (h) leads to contradictions. So we conclude that this graph in Fig.~\ref{fig:1341 and 4-point star} is not solvable.


\begin{thebibliography}{100}
\bibitem{note-diameter} The diameter of a graph is the maximum distance between any two vertices in the graph; the distance between two vertices in a graph is the length of a shortest path between the two vertices. Since we are considering only connected graphs, the diameter must be a finite integer.
\bibitem{note-odd cycles} In fact, we can also include non-bipartite graphs if we modify the definition of a layer graph by allowing edges connecting vertices in the same layer. With such a definition, all graphs can be drawn as layer graphs. But it seems that layer graph is not a convenient (although possible) way to represent non-bipartite graphs, and we are not going to consider non-bipartite graphs in our scheme.
\end{thebibliography}
 \end{document}